\documentclass[twocolumn]{aastex631}

\usepackage{amsmath} 
\usepackage{bm}      
\usepackage{graphicx}
\usepackage{booktabs} 

\newcommand{\Mpch}{\ensuremath{h^{-1}\mathrm{Mpc}}}
\newcommand{\kms}{\ensuremath{\mathrm{km~s}^{-1}}}
\newcommand{\fsigma}{f\sigma_8}
 
\shorttitle{ATLAS Growth of Structure}
\shortauthors{Dixon \& Jones}

\begin{document}

\title{The Peculiar Growth of Structure: Validating $f\sigma_8$ Measurements from TITAN Type Ia Supernovae and the Uchuu Simulations}

\correspondingauthor{Mitchell Dixon}
\email{mtdixon@hawaii.edu}

\author[0000-0003-0928-0494]{Mitchell Dixon}
\affiliation{Institute for Astronomy, University of Hawai‘i, 640 N Aohoku Place, Hilo, HI 96720, USA}

\author[0000-0002-6230-0151]{David O.~Jones}
\affiliation{Institute for Astronomy, University of Hawai‘i, 640 N Aohoku Place, Hilo, HI 96720, USA}

\author[0000-0003-4761-2197]{Nicole E.\ Drakos}
\affiliation{University of Hawai‘i at Hilo, 200 W Kawili St, Hilo, HI 96720, USA}

\author[0000-0002-1296-412X]{Bastien Carreres}
\affiliation{Department of Physics, Duke University, Durham, NC 27708, USA}

\author[0009-0004-5681-545X]{Jack W. Tweddle}
\affiliation{Astrophysics sub-Department, Department of Physics, University of Oxford, Keble Road, Oxford, OX1 3RH}

\author[0000-0002-8342-3804]{Yukei S. Murakami}
\affiliation{Department of Physics and Astronomy, Johns Hopkins University, Baltimore, MD 21218, USA}

\author[0000-0002-5221-7557]{Chris Ashall}
\affiliation{Institute for Astronomy, University of Hawai‘i, 2680 Woodlawn Dr., Honolulu, HI 96822, USA}

\author[0000-0001-5201-8374]{Dillon Brout}
\affiliation{Departments of Astronomy and Physics, Boston University, Boston MA 02215}

\author[0000-0001-6069-1139]{Thomas de Jaeger}
\affiliation{Sorbonne Universit\'{e}, CNRS, Laboratoire de Physique Nucl\'{e}aire et de Hautes Energies, 75252 Paris, France}

\author[0000-0003-3429-7845]{Aaron Do}
\affiliation{Institute of Astronomy and Kavli Institute for Cosmology, Madingley Road, Cambridge CB3 0HA, UK}

\author[0009-0003-4631-3184]{Elijah G. Marlin}
\affiliation{Departments of Astronomy and Physics, Boston University, Boston MA 02215}

\author[0000-0001-5402-4647]{David Rubin}
\affiliation{Institute for Astronomy, University of Hawai‘i, 2680 Woodlawn Dr., Honolulu, HI 96822, USA}

\author[0000-0002-8229-1731]{Stephen J. Smartt}
\affiliation{Department of Physics, University of Oxford, Denys Wilkinson Building, Keble Road, Oxford OX1 3RH, UK}
\affiliation{Astrophysics Research Centre, School of Mathematics and Physics, Queen’s University Belfast, BT7 1NN, UK}

\author[0000-0001-9535-3199]{Ken W. Smith}
\affiliation{Department of Physics, University of Oxford, Denys Wilkinson Building, Keble Road, Oxford OX1 3RH, UK}
\affiliation{Astrophysics Research Centre, School of Mathematics and Physics, Queen’s University Belfast, BT7 1NN, UK}

\begin{abstract}
The growth rate of cosmic structure, parameterized by $f\sigma_8$, is a fundamental test of $\Lambda$CDM and general relativity. Using Type~Ia supernova (SN\,Ia) peculiar velocities in conjunction with galaxy redshift surveys may be one of the most precise pathways to measuring $f\sigma_8$ in the local Universe, yet existing analyses have not quantified its systematic uncertainties. Here, we present an end-to-end simulated $f\sigma_8$ measurement tailored to the Type\,Ia Supernova Trove from ATLAS in the Nearby Universe (TITAN) survey, using $\sim$2000 simulated SNe\,Ia at $z < 0.067$. We use the Uchuu $N$-body simulations to generate mock galaxy catalogs and SN\,Ia simulations with realistic, correlated peculiar velocities, and use these catalogs to reconstruct density fields that replicate the 2M$++$ redshift survey. Using a modified forward likelihood framework across eight mock realizations, we recover $\langle f\sigma_8 \rangle = 0.429 \pm 0.038$ ($\sigma_{\rm stat} = 0.030$, $\sigma_{\rm sys} = 0.023$), consistent with the Uchuu simulation input $f\sigma_8 = 0.428$ to within $0.1\%$. The mock‑to‑mock scatter of $0.031$ is consistent with our uncertainties, highlighting the reliability of our error estimates. Our measurement is dominated by the statistical uncertainty, with approximately equal contributions from SN\,Ia ($\sigma_{\rm sys}^{\rm SN}=0.017$) and density reconstruction ($\sigma_{\rm sys}^{\rm recon}=0.016$) systematic uncertainties. The assumed intrinsic scatter model is the largest single systematic contribution, and a different model choice can shift $f\sigma_8$ to lower values, largely driven by red SNe with a skewed color distribution. Our analysis provides the first systematic uncertainty budget for the ``reconstruction-and-scaling'' method of measuring $f\sigma_8$ with SNe\,Ia, and demonstrates that SNe\,Ia are a competitive probe of the growth of structure in the local Universe.
\end{abstract}


\section{Introduction} \label{sec:intro}

The $\Lambda$CDM model has emerged as the standard framework for cosmology, successfully describing a wide range of observations from the cosmic microwave background (CMB) to the large-scale structure of the local Universe (\citealp{Planck2018, Troster_2020, Heymans_2021, Abbott_2020}). However, precision measurements of cosmological parameters have revealed significant tensions between early and late Universe probes (see \citealp{Verde_2019, Di_Valentino_2021} for recent reviews). The most prominent is the Hubble tension: a $\sim5$--$6\sigma$ discrepancy between the Hubble constant $H_0$ derived from the CMB \citep{Planck2018} and the local distance ladder (\citealp{Riess_2022}, albeit with some studies finding smaller tensions, e.g., \citealp{Freedman_2021}). Beyond $H_0$, a similar tension exists in measurements of the growth rate of cosmic structure. Parameterized by $f\sigma_8$, where $f$ is the growth rate and $\sigma_8$ is the root mean square (RMS) of matter fluctuations on $8\,\Mpch$ scales, the growth rate of structure provides a fundamental test of gravity and $\Lambda$CDM. Local weak lensing and galaxy clustering surveys \citep{Heymans_2021, Abbott_2022, Karim_2025} consistently favor a lower $\sigma_8$, and find $\sim 3\sigma$ tension with results from the CMB under the assumption of $\Lambda$CDM.

Peculiar velocities, deviations from the smooth Hubble flow caused by the gravitational pull of local matter overdensities, offer a direct and independent probe of this growth rate (\citealp{Peebles_1980, Davies_2011,Hudson_2012,Turnbull_2012, Huterer_2017, Kim_2019, Adama_2020, Boruah_2020, Stahl_2021, Do_2025}). 
Type Ia supernovae (SNe Ia) can be used as precise peculiar velocity probes by comparing the redshifts of their host galaxies with the redshifts inferred from their luminosity distances \citep{Miller_1992, Riess_1997, Gordon_2007, Weyant_2011}. SN Ia standardized luminosities provide distance estimates with 5--7\% precision \citep{Scolnic_2018, 2022Brout}, significantly better than the $\sim$20\% accuracy of galaxy-based distance methods such as the Tully--Fisher relation or the Fundamental Plane \citep{Tully1977,Dressler87,Djorgovski87}. However, because the dynamical influence of peculiar velocities is most significant at low redshifts ($z < 0.1$), previous studies have historically been limited to samples of only a few hundred SNe~Ia \citep{Turnbull_2012, Huterer_2017, Boruah_2020, Stahl_2021}. 

However, recent surveys such as the Foundation Supernova Survey \citep{Foley18,Jones19} have helped to revitalize the low-$z$ ($z \lesssim 0.1$) SN\,Ia sample, and all-sky surveys such as ZTF \citep{Bellm19} are now adding thousands of new low-$z$ SNe\,Ia to the Hubble diagram \citep{Dhawan22,Rigault25}. Complementing this new wealth of low-$z$ data, the Asteroid Terrestrial-impact Last Alert System (ATLAS; \citealp{Tonry_2018}) provides an all-sky, high-cadence survey that has discovered thousands of SNe Ia \citep{Smith_2020_ATLAS,Murakami_2026} and produced high quality lightcurves \citep{Marlin_2025}. The Type Ia Supernova Trove from ATLAS in the Nearby Universe (TITAN) compilation (Murakami et al.\ in prep, Tweddle et al.\ in prep, \citealp{Marlin_2025}) is currently the largest low-$z$ sample available for peculiar velocity studies (although ZTF may soon surpass them following improvements in their calibration; \citealp{Lacroix25,Kenworthy25}).

As sample sizes increase, the limiting factor in measuring $f\sigma_8$ shifts from statistical to systematic uncertainties. For SNe~Ia, non-uniform sky coverage, photometric calibration error, dispersion in the Milky Way dust reddening law, and correlations between host galaxy properties and SN Ia luminosity can all bias cosmological parameters \citep[e.g.,][]{Brout_2019, Jones19, 2022Brout, vincenzi2024dark}. Measurements of $f\sigma_8$ also appear particularly sensitive to the choice of SN\,Ia intrinsic scatter model \citep{Carreres_2025}. In the case of the ``reconstruction-and-scaling'' approach \citep{Turnbull_2012, Boruah_2020, Stahl_2021}, where observed SN\,Ia peculiar velocities are compared to the matter density field derived from a galaxy redshift survey, additional uncertainties arise from the reconstruction itself. This includes the choice of smoothing scale, the survey volume, and the completeness of the galaxy catalog \citep{Hollinger21,Turner_2023,Hollinger24}. To fully exploit the statistical power of modern SN samples, both SN\,Ia and reconstruction systematics must be properly understood and propagated.

However, existing measurements of $f\sigma_8$ using SNe\,Ia typically only include statistical errors in their analysis \citep{Boruah_2020,Stahl_2021}. Moreover, two recent papers find evidence that density reconstruction-based measurements of $f\sigma_8$ may have underestimated uncertainties and significant parameter biases \citep{Turner_2023,Blake24}.  To provide robust constraints on $f\sigma_8$, we therefore require validation of this method via simulations, and a full accounting of the uncertainties attributable to SN\,Ia and density reconstruction systematics.

Here, we use the Uchuu N-body simulations \citep{Ishiyama_2021}, the associated UniverseMachine galaxy catalogs \citep{Behroozi_2019, Aung_2023}, and SN\,Ia simulations developed for TITAN (Tweddle et al.\ in prep.) to create an end-to-end simulated reconstruction-and-scaling measurement of $f\sigma_8$.  We adopt the method of \citet{Carreres_2025} to map density fluctuations in the Uchuu simulations to the peculiar velocities used to generate the resulting SN\,Ia simulations. We use this measurement to evaluate the effectiveness of both the method itself, and the sizes of the associated systematic uncertainties in a TITAN SN\,Ia $f\sigma_8$ measurement.

This paper is organized as follows. We describe our simulations and mock catalogs in Section~\ref{sec:simulations}, our velocity field reconstruction in Section~\ref{sec:reconstruction}, and our forward likelihood framework and systematic covariance treatment in Section~\ref{sec:fsigma8}. Results are presented in Section~\ref{sec:results}, limitations and implications for the upcoming TITAN observational analysis are discussed in Section~\ref{sec:discussion}, before we conclude in Section~\ref{sec:conclusions}.

\section{Simulations and Mock Catalogs}\label{sec:simulations}

\subsection{Uchuu N-body Simulation}\label{subsec:uchuu}
We use the Uchuu N-body simulation \citep{Ishiyama_2021} as the foundation of our SN simulations and density reconstructions. Uchuu contains $12800^3$ dark matter particles each with mass $m_p = 3.27 \times 10^8 \, M_\odot$ in a $2.0 \, \mathrm{Gpc} \, h^{-1}$ box, initialized at $z = 127$ using \citet{Planck_2015} cosmology; $\Omega_m = 0.3089$, $\Omega_\Lambda = 0.6911$, $h = 0.6774$, and $\sigma_{8,m} = 0.8159$. The UniverseMachine galaxy catalog \citep{Behroozi_2019, Aung_2023} is used to populate halos with galaxies using abundance matching and empirical scaling relations that match observed stellar mass functions and star formation rates. To account for cosmic variance, we extract eight independent subvolumes of side length $400\,\Mpch$ from the full Uchuu simulated box, and transform each into a Galactic coordinate system.



\subsection{Simulated ATLAS Type Ia Supernovae}\label{subsec:mock_sne}

The TITAN compilation has recently assembled a sample of $\sim$8000 spectroscopically confirmed SNe\,Ia, more than $\sim$5000 of which have host-galaxy redshifts (Murakami et al. in prep, Tweddle et al. in prep). Its distance measurements have a Hubble diagram scatter of $\sim$0.17~mag and a systematic uncertainty in the photometric calibration of just $\sim$5--10~mmag \citep{Marlin_2025}. At $z < 0.067$ --- corresponding to the redshift limit of the 2M$++$ density reconstruction \citep{Carrick_2015} --- the ATLAS sample comprises $\sim$2,000 SNe\,Ia, after cosmology grade cuts (Murakami et al. in prep.).

Here, we create SN simulations based on the TITAN sample by using the \textsc{SNANA} simulation package \citep{Kessler_2009} within the {\tt Pippin} pipeline \citep{Hinton_2020}. SNANA is a simulation and analysis software tool that performs realistic SN\,Ia survey simulations, SN\,Ia light-curve fitting, and cosmological parameter estimation. SNANA simulations are designed to replicate realistic survey properties, including cadence, filter transmission, and noise (see \citealp{Kessler_2019} for a detailed overview).  Our full simulation parameters will be presented in Tweddle et al.\ (in prep), and reliably model the real ATLAS observing strategy, redshift distribution, filter transmission functions, image zeropoints, and the resulting signal-to-noise ratios (S/N) of the data.  We use SALT3 \citep{Kenworthy_2021} as our source model, adopting the version that was re-calibrated by the Dark Energy Survey (DES) team \citep{Taylor_2023}.  

To generate the intrinsic dispersion of our simulated SNe\,Ia --- the residual scatter in standardized luminosities beyond what measurement uncertainties can account for --- we adopt the \citet[][hereafter BS21]{brout2021dust} scatter model, with parameters drawn from \citet{Popovic_2023} that closely match the TITAN data.  The BS21 model assumes that scatter from diverse dust reddening laws dominates SN scatter, and has been empirically shown to be the best match to SN\,Ia Hubble diagram data \citep[e.g.,][]{2022Brout,vincenzi2024dark}. We also include two additional scatter models as systematic variants: the \citet[][hereafter G10]{G10} model, in which the scatter is $\sim25\%$ chromatic and $\sim$75$\%$ achromatic, and the \citet[][hereafter C11]{Chotard11} model, in which most scatter is wavelength-dependent. The simulated $x_1$ and $c$ distributions are adopted from the DES five-year analysis \citep{vincenzi2024dark}, and are largely indistinguishable except that they have a slightly longer tail at redder colors than the BS21 $c$ distribution; these distributions will be re-tuned to match the TITAN data in Tweddle et al.\ (in prep.), and this adjustment may slightly reduce intrinsic scatter systematics (see Section \ref{sec:scatter}).

\begin{figure}
    \centering
    \includegraphics[width=1.0\linewidth]{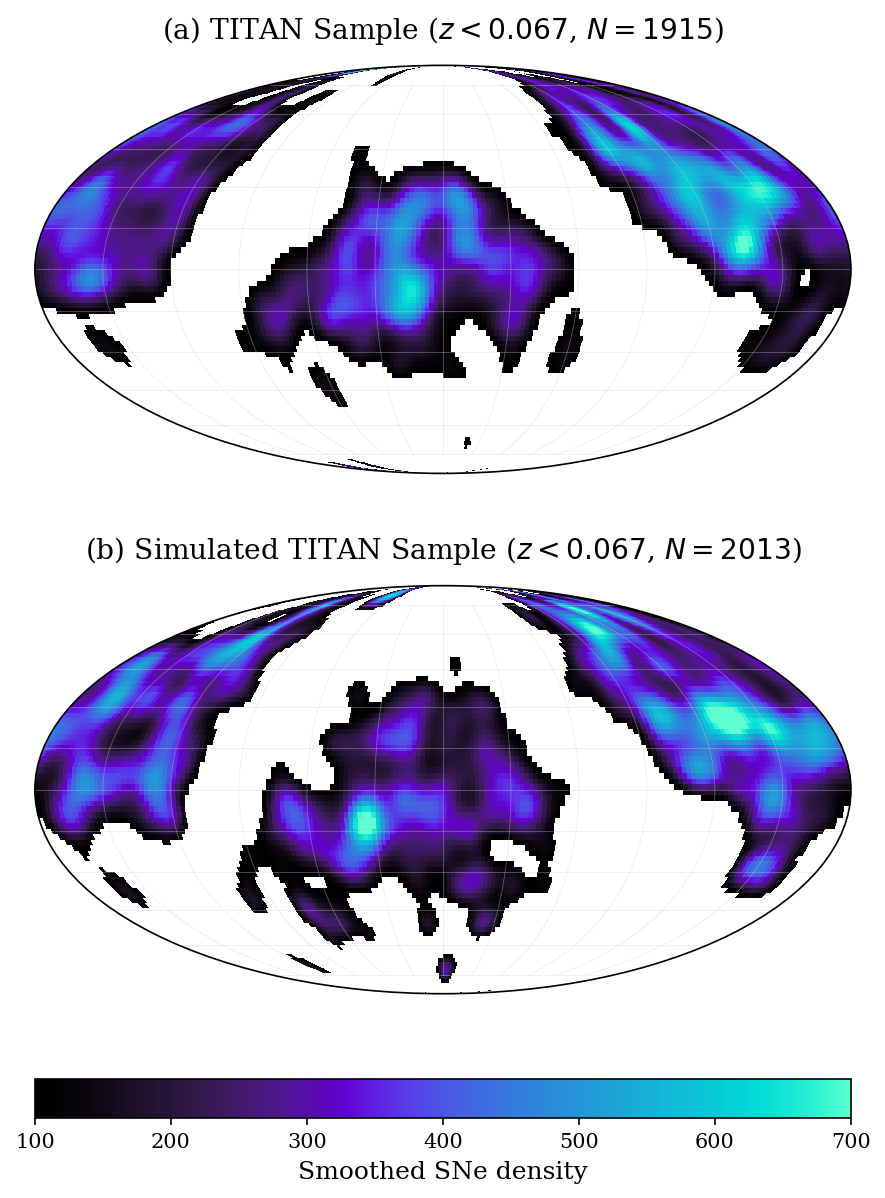}
    \caption{Sky distribution comparison between the observed TITAN sample (top) and an example simulated ATLAS mock catalog (bottom). Both panels show SNe~Ia with $z <0.067$ in an equatorial Mollweide projection. The colorbar highlights the smoothed SNe density, using the same scale for the observed and simulated samples.}
    \label{fig:atlas_simlib}
\end{figure}

\begin{figure}
    \centering
    \includegraphics[width=1.0\linewidth]{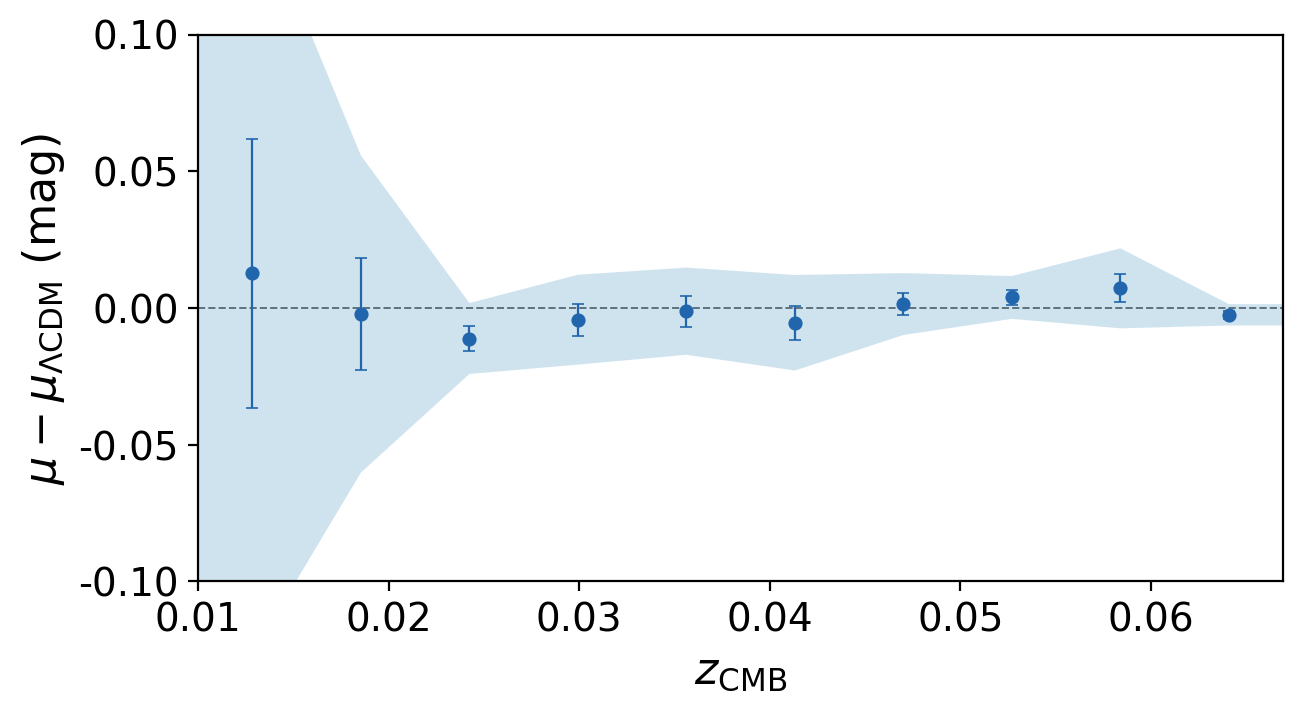}
    \caption{Hubble diagram residuals for the eight baseline Uchuu mock realizations, each containing ${\sim}2000$ SNe~Ia ($z_{\rm CMB}<0.067$). Residuals are computed from the bias-corrected distance moduli as $\mu-\mu_{\Lambda{\rm CDM}}$, using a fixed flat $\Lambda$CDM reference cosmology. The blue points show the mean binned residual across the mocks, and the shaded band highlights the mock-to-mock scatter.}
    \label{fig:atlas_sim_HD}
\end{figure}

To simulate host-galaxy properties and peculiar velocities that faithfully follow the matter distribution in the Uchuu simulations, we use the method of \citet{Carreres_2025} to incorporate these parameters in SNANA's host-galaxy library ({\tt HOSTLIB}). We then construct a ``simulation library'' ({\tt SIMLIB}) of randomly-sampled right ascensions ($\alpha$) and declinations ($\delta$) using galaxy positions from the Uchuu UniverseMachine catalog \citep{Ishiyama_2021, Aung_2023}, and match galaxy properties to each SN based on those coordinates. These properties include peculiar velocities and stellar masses. We also take host-galaxy magnitudes and S\'ersic profiles from the OpenUniverse LSST--Roman simulations \citep{OpenUniverse_2025}, interpolated in $(\log M_*, \log{\rm SFR})$ space, which model the host contribution to the SN\,Ia flux and the placement of each SN within its host.

Because ATLAS has fewer classified SNe\,Ia in the Southern sky and $f\sigma_8$ measurements are sensitive to correlations across large spatial scales, we also weight the distribution of galaxy coordinates to reproduce the observed ATLAS ($\alpha$, $\delta$) sky distribution using a HEALPix density map constructed from the real TITAN SN Ia sample (Figure~\ref{fig:atlas_simlib}). 

Next, we create eight independent ATLAS mock catalogs, one from each Uchuu subvolume, allowing us both to test our analysis and reconstruction pipeline using multiple density fields, and to understand the effects of cosmic variance. We then generate two SN\,Ia simulations for each mock: one that aims to match the number of SNe\,Ia in the real data, and a second (``{{\tt BiasCor}}") simulation with $\gtrsim$50,000 events that we use to correct for the distance bias ($\Delta\mu = \mu_{\rm obs} - \mu_{\rm sim}$) as a function of redshift following standard procedures used for real SN\,Ia analyses. The Hubble diagram for all eight mock realizations is shown in Figure~\ref{fig:atlas_sim_HD}.

\section{Velocity Field Reconstruction}\label{sec:reconstruction}

In this section, we reconstruct the density and velocity fields from the Uchuu mock galaxy catalogs, following an approach designed to closely match the 2M$++$ reconstruction of \citet{Carrick_2015}.

\subsection{Peculiar Velocities from Linear Perturbation Theory}

Peculiar velocities are induced by gravitational acceleration arising from the underlying matter distribution. In the linear regime of structure formation, the relationship between the velocity field, $\mathbf{v}(\mathbf{r})$, and the matter density contrast, $\delta(\mathbf{r}) \equiv (\rho({\mathbf r}) - \overline{\rho})/\overline{\rho}$, is given by:
\begin{equation}
\mathbf{v}(\mathbf{r}) = \frac{H_0 f}{4\pi} \int d^3\mathbf{r}'\, \delta(\mathbf{r}') \frac{\mathbf{r}' - \mathbf{r}}{|\mathbf{r}' - \mathbf{r}|^3},
\label{eq:linear_velocity}
\end{equation}
where $H_0$ is the Hubble constant, and $f$ is the growth rate ($f \approx \Omega_m^{0.55}$ for $\Lambda$CDM). 

However, observational surveys map the distribution of luminous galaxies, while the induced peculiar velocities are driven by the total matter distribution (which is dark matter-dominated). We therefore adopt a linear bias model $\delta_g = b\delta$, where $b$ is the linear galaxy bias and $\delta_g$ is the galaxy overdensity. Equation \ref{eq:linear_velocity} then becomes:
\begin{equation}
\mathbf{v}(\mathbf{r}) = \frac{H_0 \beta}{4\pi} \int d^3\mathbf{r}'\, \delta_g(\mathbf{r}') \frac{\mathbf{r}' - \mathbf{r}}{|\mathbf{r}' - \mathbf{r}|^3} + \mathbf{V}_{\text{ext}},
\label{eq:practical_reconstruction}
\end{equation}

\noindent where $\beta \equiv f/b$ is the scaling amplitude between the observed galaxy density and the velocity field. We also include an external bulk flow term ($\mathbf{V}_{\text{ext}}$), to account for velocity contributions from structures beyond our reconstruction volume. In this analysis, we obtain $f\sigma_8$ using the relation $f\sigma_8 = \beta\,\sigma_{8,g}$, where $\beta$ is the best-fit value and $\sigma_{8,g}$ is the RMS of galaxy density fluctuations measured on $8\,\Mpch$ scales from the galaxy density field. 

\subsection{Density Reconstruction}\label{subsec:density_recon}

The density field reconstruction of \citet{Carrick_2015} is derived from the 2M$++$ galaxy redshift catalog \citep{Lavaux_2011}.  2M$++$ combines the Two-Micron All-Sky Redshift Survey (2MRS; \citealp{Huchra_2012}), the 6dF Galaxy Redshift Survey (6dFGS; \citealp{Jones09}), and the Sloan Digital Sky Survey Data Release 7 (SDSS DR7; \citealp{Abazajian_2009}) to provide a nearly all-sky map of the local galaxy distribution. In regions covered only by 2MRS the redshift sample is magnitude-limited to $K \leq 11.5$, extending to $\sim125~\Mpch$, while the addition of 6dFGS and SDSS redshifts deepens the catalog to $K \leq 12.5$, which reaches $\sim 200~\Mpch$.  Below, we describe the procedures by which we apply and correct for the 2M$++$ selection criteria to construct a density field using the Uchuu simulations.


\subsubsection{Assigning $K$-band Magnitudes}\label{subsec:abundance_matching}

Replicating the 2M$++$ selection requires $K$-band magnitudes for each galaxy, so that we can apply an apparent magnitude cut of $K < 12.5$ and restrict our science region to a radius of $200\,\Mpch$. The OpenUniverse host-galaxy magnitudes we use for the SN simulations (Section~\ref{subsec:mock_sne}) are unsuitable here. They saturate in the $K$ band at approximately 13--14~AB mag ($\sim$11.5--12.5~mag in the Vega system), close to the 2M$++$ detection limit, leaving the bright, nearby galaxies in our mock catalogs with unreliable magnitudes. Hence, we choose to assign $K$-band absolute magnitudes through abundance matching to the virial masses ($M_{\rm vir}$) of each halo, before applying the $K < 12.5$ cut to our sample, as described below.

We sample the $K$-band Schechter luminosity function \citep{Schechter_1976} with the parameters used by \citet{Carrick_2015}: $M^{*} - 5\log_{10}h = -23.25$, $\alpha = -0.85$, and $\phi^{*} = 0.0108\,h^{3}\,\mathrm{Mpc}^{-3}$, determined by  \citet{Cole_2001}. We integrate the luminosity function over the effective volume of our mock catalog to determine the total number of galaxies predicted. We then draw this number of random magnitudes from the luminosity function via inverse cumulative distribution function sampling. These magnitudes are ranked and matched to halos sorted by $M_{\mathrm{vir}}$, where the most massive halo receives the brightest magnitude up to a faint-end threshold of $M_{K} = -20$~mag (Vega). Finally, we convert the assigned absolute magnitudes to apparent magnitudes using the distance moduli ($\mu$). Our resulting sample contains $\sim$85,000 galaxies within $r < 200\,\Mpch$, compared with the $\sim$70,000 in the 2M$++$ catalogue \citep{Lavaux_2011}. We simplify the 2M$++$ selection by applying this cut uniformly across the sky rather than reproducing the shallower $K \leq 11.5$ limit of the 2MRS-only regions. This retains nearby galaxies that 2M$++$ excludes and largely accounts for the difference.

\begin{figure}
    \centering
    \includegraphics[width=0.8\linewidth]{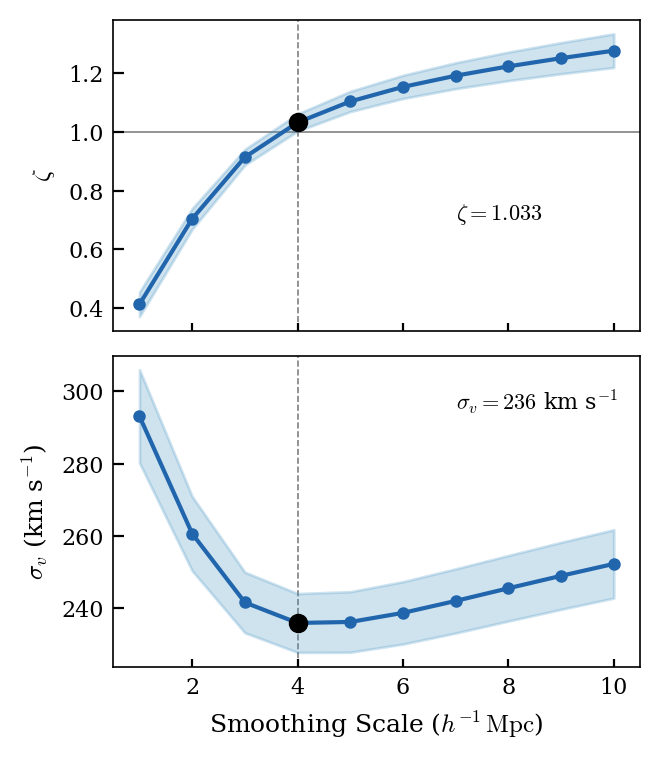}
    \caption{{\bf Top:} Regression slope ($\zeta$) between the reconstructed and true radial velocities as a function of smoothing scale across the eight Uchuu mocks. The horizontal line denotes an unbiased reconstruction ($\zeta = 1$). {\bf Bottom:} The residual velocity dispersion $\sigma_v$. The vertical dashed line highlights our chosen optimal scale of $R_s = 4~h^{-1}\mathrm{Mpc}$, which minimizes $\sigma_v$ while keeping the reconstruction slope consistent with unity ($\zeta = 1.033 \pm 0.029$).}
    \label{fig:smoothing}
\end{figure}

\subsubsection{Selection Weights}\label{subsec:weights}

Flux-limited galaxy surveys become increasingly incomplete at large distances, where only the brightest galaxies exceed the apparent magnitude threshold. To correct for this incompleteness in our mock catalogs, we assign luminosity-based selection weights to each galaxy, which increase a galaxy's contribution when constructing the density field, following the approach of \citet{Carrick_2015} and \citet{Hollinger24}. The luminosity-weighted selection weight at distance $r$ is:

\begin{equation}
w^{L}(r) = \frac{\int_{L_{\min}}^{\infty} L \, \Phi(L) \, dL}{\int_{L(r)}^{\infty} L \, \Phi(L) \, dL} 
\end{equation}

\begin{figure*}
    \centering
    \includegraphics[width=1\linewidth]{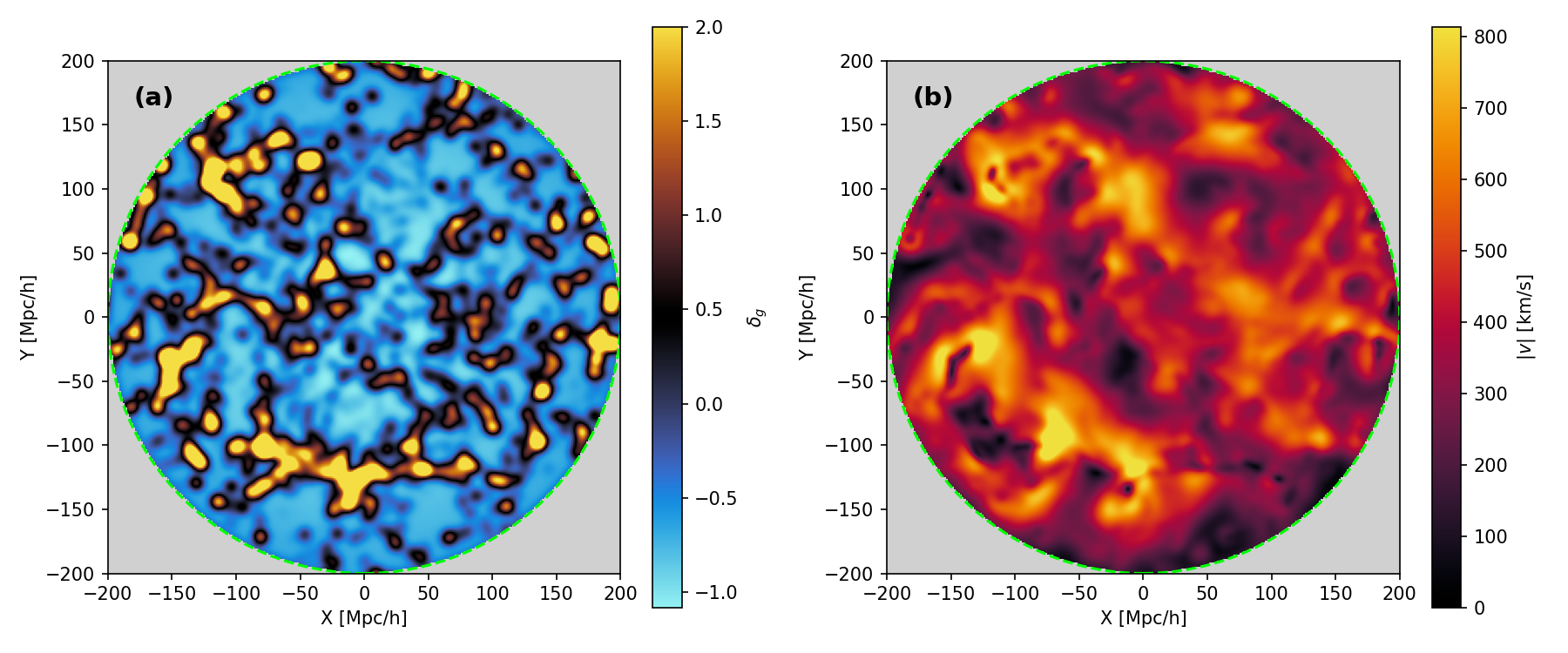}
    \caption{Example reconstructed fields from a single Uchuu mock galaxy catalog in Galactic Cartesian coordinates, including \textbf{(a)} the luminosity-weighted galaxy density contrast ($\delta_g$) and \textbf{(b)} the velocity magnitude ($|v|$) in \kms. Both panels show a $Z=0$ slice with $4~\Mpch$ Gaussian smoothing. The dashed green circle marks $r = 200~\Mpch$, the boundary within which we restrict our analysis to match the effective volume of the 2M$++$ reconstruction \citep{Carrick_2015}.}
    \label{fig:mock_00}
\end{figure*}

\begin{figure*}
    \centering
    \includegraphics[width=0.9\linewidth]{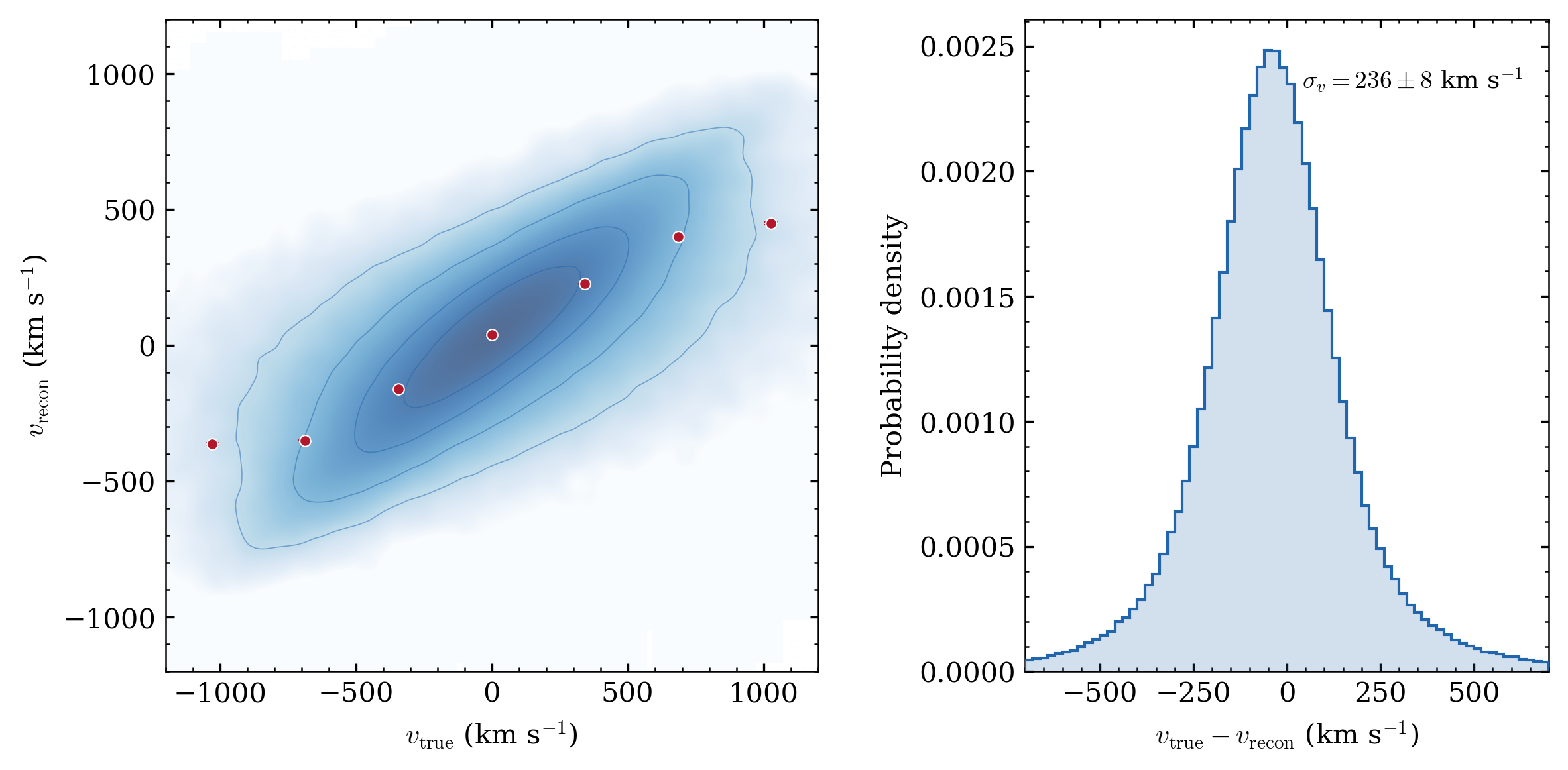}
    \caption{Velocity recovery from the eight Uchuu mock reconstructions. \textbf{Left:} reconstructed versus true radial velocities for galaxies within $200~\Mpch$, with binned medians shown in red. \textbf{Right:} velocity residual distribution, with $\sigma_v = 236 \pm 8\,\kms$ and a median offset of $-37\,\kms$ across the combined mocks.}
    \label{fig:velocity_recovery}
\end{figure*}

\noindent where $L_{\min}$ is the faint-end threshold corresponding to $M_K = -20$, and $L(r)$ is the minimum luminosity detectable at $r$, given the $K < 12.5$ magnitude limit. We treat this limit as homogeneous across the sky, so the selection function depends only on distance. The weights remain close to unity at small distances and rise smoothly to $w \approx 7$ within our volume ($r = 200~\Mpch$), consistent with the 2M$++$ luminosity weights obtained in \citet{Carrick_2015}.

\subsubsection{Bias Corrections}\label{subsec:bias_corrections}

While selection weights help correct for distance-dependent incompleteness in the galaxy number counts, an additional bias correction $F(r)$ is needed to normalize the density contrast to a uniform effective galaxy bias, since the luminosity threshold of the surviving galaxies increases with distance and more luminous galaxies are more biased tracers of the underlying matter. We follow \citet{Carrick_2015}, who adopt the luminosity-dependent bias relation of \citet{Westover_2007}:

\begin{equation}
  F(r) = \frac{b}{b^{*}} = 0.73 + 0.24\, L/L^{*},
\end{equation}

\noindent where $L/L^{*}$ is the mean luminosity of observable galaxies in units of the characteristic luminosity $L^{*}$, and $b^{*}$ is the bias of an $L^{*}$ galaxy. At small distances $L/L^{*}$ is low and $F(r) \sim 1$, and both rise with increasing distance. We measure $L/L^{*}$ in radial shells to compute $F(r)$ from the flux-limited catalog, interpolate onto the 3D grid, and then divide the density field by $F(r)$ to normalize it to a uniform effective bias.

\subsubsection{Reconstructed Density and Velocities}\label{subsec:recon_results}

Putting this all together, we use our catalog of $K < 12.5$ galaxies to construct a 3D density field on a uniform grid of $257^3$ cells.  Because the 2M$++$ flux limit effectively restricts our science region to a radius of $200\,\Mpch$, we compute the density field over a box of $400\,\Mpch$ per side ($\sim 1.56\,\Mpch$ cell size), centered on the observer in Galactic Cartesian coordinates $(X, Y, Z)$. 

We assign galaxies to grid cells using the Cloud-In-Cell (CIC) mass-assignment scheme implemented in {\tt Pylians} \citep{Pylians}, weighting each by its selection weight $w^L(r)$ (Section~\ref{subsec:weights}), and normalize the resulting field to the mean density within the sphere to obtain the overdensity $\delta_g = \rho/\bar{\rho} - 1$. Outside the $200\,\Mpch$ radius but within our cube, we set $\delta_g = 0$ as a mean-density boundary condition, consistent with \citet{Carrick_2015}. We then divide by the radially dependent bias correction $F(r)$ to normalize the density field to a uniform effective galaxy bias across the survey volume.

Lastly, we smooth the density field of each mock with a Gaussian kernel of radius $R_s = 4\,\Mpch$. As shown in Figure~\ref{fig:smoothing}, this smoothing scale minimizes the residual velocity dispersion $\sigma_v$. Smaller scales fail to suppress non-linearities, while larger scales over-smooth local density gradients; both increase the velocity residuals. This $4\,\Mpch$ scale matches the optimal choice found by \citet{Berlind_2000}, \citet{Carrick_2015}, and \citet{Hollinger21}.

We quantify the reconstruction quality using the regression slope $\zeta$ between the true and reconstructed radial velocities, finding $\zeta = 1.033 \pm 0.029$ across the eight mocks, consistent with unbiased recovery. Figure~\ref{fig:mock_00} shows the density contrast ($\delta_g$) and velocity magnitudes ($|v|$) for a single mock in the $XY$ plane at $Z = 0$, while Figure~\ref{fig:velocity_recovery} shows the velocity recovery across all eight mocks, with a residual scatter of $\sigma_v = 236 \pm 8\,\kms$. We also measure $\sigma_{8,g} = 1.06 \pm 0.03$ from the RMS of the reconstructed density field smoothed with an $8\,\Mpch$ top-hat filter. While the reconstruction extends to $200\,\Mpch$, we evaluate $\sigma_{8,g}$ within $190\,\Mpch$ to avoid edge effects near the boundary.

\section{Measuring the Growth Rate of Structure}\label{sec:fsigma8}

Several methods exist for measuring $f\sigma_8$ using peculiar velocities, including maximum-likelihood fitting, power spectrum analysis, and velocity field reconstruction (see \citealt{Turner_2024} for a review). In this work, we use the reconstruction-and-scaling approach \citep{Boruah_2020, Stahl_2021}. Below, we describe our SN\,Ia distances (Section~\ref{sec:distances}), our adapted forward likelihood framework (Section~\ref{sec:likelihood}), how we incorporate systematic covariances (Section~\ref{sec:likelihood_sys}), and our choice of systematic uncertainties in measuring $f\sigma_{8}$ (Section~\ref{sec:systematics}).

\subsection{Measuring SN Ia Distances}
\label{sec:distances}

We measure the SN\,Ia light-curve parameters $x_0$, $x_1$, and $c$ (and their associated uncertainties and covariances) by fitting our simulated light curves with the SALT3 model. The $x_0$ parameter is the amplitude, related to the peak rest-frame $B$-band magnitude by $m_B = -2.5\log_{10}(x_0) + {\rm constant}$; $x_1$ is the first principal component of variation, which correlates with light-curve shape; and $c$ is a color parameter that describes both dust attenuation and intrinsic SN\,Ia color. We apply standard light-curve selection criteria \citep[e.g.,][]{2022Brout,vincenzi2024dark}, requiring $-3 < x_1 < 3$, $-0.3 < c < 0.3$, $\sigma_{x_1} < 1$, the uncertainty on time of maximum light $\sigma_{t_0} < 1~{\rm days}$, and $\sigma_c < 0.1~{\rm mag}$.

The SALT3 light-curve parameters are converted into distances using the modified Tripp equation \citep{1998A&A...331..815T},
\begin{equation}
\mu = m_B + \alpha_{\rm SN} x_1 - \beta_{\rm SN} c + \gamma_{\rm SN} G_{\rm host} - \mathcal{M} - \delta_{\rm bias},
\label{eq:tripp}
\end{equation}
where $G_{\rm host} = +1/2$ for high-mass  ($\log_{10}(M_{\ast}/M_{\odot}) > 10$) and $-1/2$ for low-mass host galaxies ($\log_{10}(M_{\ast}/M_{\odot}) < 10$). The parameter $\gamma_{\rm SN}$ is the size of the resulting host-galaxy mass step \citep{kelly2010,Lampeitl_2010,Sullivan_2010}, and a $0.1$~mag offset in absolute rest-frame $B$-band magnitude between the two populations after light-curve corrections is adopted in our simulations. $\mathcal{M}$ is the mean SN~Ia absolute magnitude (degenerate with $H_0$) while the $\delta_{\rm bias}$ term corrects for observational biases and selection effects in the measured distances. These corrections are derived from a large ``{\tt BiasCor}'' simulation in which the offset between simulated and recovered SN\,Ia distances is estimated as a function of redshift and applied to the ``data'' simulation. Systematic uncertainties in this correction are included in the covariance matrix (Section~\ref{sec:systematics}).

The nuisance parameters $\alpha_{\rm SN}$, $\beta_{\rm SN}$, $\gamma_{\rm SN}$, and $\mathcal{M}$ are estimated simultaneously alongside redshift-dependent binned distances following \citet{Marriner11}, as implemented in the BEAMS with Bias Corrections (BBC) method of \citet{Kessler_2017}. This procedure breaks the degeneracy between nuisance parameter estimation and the cosmological model. Our distance uncertainties do not include peculiar velocity uncertainty, as this is modeled by the forward likelihood (Section~\ref{sec:likelihood}), but do include a lensing uncertainty of $0.055z$ \citep{Jonsson10}, which is redshift dependent. The nominal Hubble diagram scatter after light-curve corrections is $\sim 0.17$~mag for all intrinsic scatter models.

\subsection{The Forward Likelihood Method}
\label{sec:likelihood}

To measure $f\sigma_8$ we employ the forward likelihood method of \citet{Boruah_2020} and \citet{Stahl_2021}, rather than estimating peculiar velocities directly by inverting the distance-redshift relation. By integrating over comoving distance along the line of sight, weighted by the local density field, this approach mitigates inhomogeneous Malmquist bias, wherein SNe~Ia are preferentially sampled from higher-density regions. Here, we first summarize the forward likelihood for individual SNe, then present our extended framework to incorporate systematic covariance. 

The predicted redshift for the $i$th SN, $z_{{\rm pred},i}$, combines the cosmological redshift and peculiar velocity as follows:
\begin{equation}
(1 + z_{{\rm pred},i}) = (1 + z_{{\rm cos},i})\left(1 + \frac{v_{{\rm pec},i}}{c}\right),
\label{eq:redshift_relation}
\end{equation}
\begin{equation}
v_{{\rm pec},i} = (\beta\,\mathbf{v}_i + \mathbf{V}_{\rm ext}) \cdot \hat{\mathbf{r}}_i,
\label{eq:vpec}
\end{equation}
where $\hat{\mathbf{r}}_i$ is a unit vector along the line of sight, $\mathbf{v}_i \equiv \mathbf{v}(\mathbf{r})$ is the reconstructed 3D velocity field evaluated at the comoving position $\mathbf{r} = r\hat{\mathbf{r}}_i$ and scaled by $\beta$, and $\mathbf{V}_{\rm ext}$ accounts for bulk flow contributions from beyond the reconstruction volume. 

The likelihood of observing redshift $z_i$ at a given position $\mathbf{r}$ is defined as
\begin{equation}
P(z_i \mid \mathbf{r}, \beta, \mathbf{V}_{\rm ext}, \sigma_v) = \frac{1}{\sqrt{2\pi\sigma_v^2}} \exp\left\{-\frac{\Delta v_i^2}{2\sigma_v^2}\right\},
\label{eq:likelihood}
\end{equation}
where $\Delta v_i \equiv cz_i - cz_{\rm pred} (\mathbf{r}, \beta, \mathbf{V}_{\rm ext})$ is the velocity residual, with $cz_{\rm pred}$  defined by Equations~\ref{eq:redshift_relation} and \ref{eq:vpec}, and $\sigma_v$ is a free parameter representing the velocity scatter not captured by the reconstruction. 

\begin{figure*}
\centering
\includegraphics[width=\linewidth]{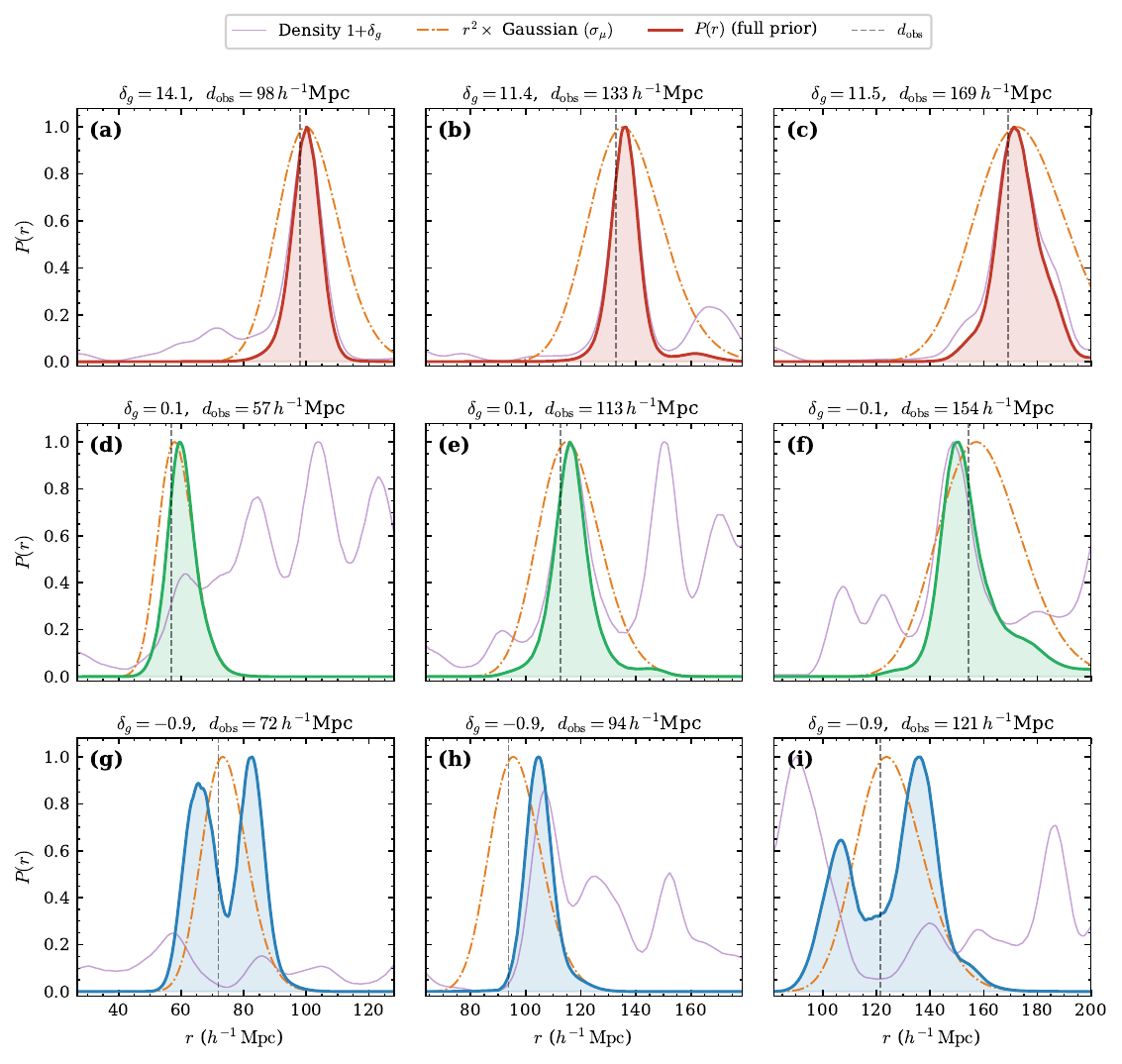}
\caption{Line of sight distance prior $P(r)$ (Equation~\ref{eq:prior}) for nine example SNe\,Ia from a single Uchuu mock, decomposed into the reconstructed density $1+\delta_g(r)$ (purple), the volume-weighted distance likelihood (orange dashed), and the full distribution (red), with the observed distance $d_{\rm obs}$ also shown (grey dashed). Local density decreases from top to bottom and columns show increasing distance. In overdense regions the density structure sharpens $P(r)$, while in voids the suppressed density produces multimodal distributions that a simple Gaussian approximation cannot model adequately.}
\label{fig:los_posterior}
\end{figure*}

Since the angular position of each SN is fixed by its observed coordinates, the 3D position reduces to a radial distance along the line of sight. Marginalizing over the unknown distance yields
\begin{equation}
P(z_i \mid \beta, \mathbf{V}_{\rm ext}, \sigma_v) = \int_0^R dr\, P(z_i \mid \mathbf{r}, \beta, \mathbf{V}_{\rm ext}, \sigma_v)\, P(r),
\label{eq:los_integral}
\end{equation}
where $R = 200\,h^{-1}\,{\rm Mpc}$ is the boundary of the 2M$++$ reconstruction \citep{Carrick_2015}. $P(r)$ is the distance prior and consists of three components:
\begin{equation}
P(r) \propto \underbrace{r^2}_{\substack{\rm volume\\\rm weighting}} \;\times\; \underbrace{\exp\Bigl[-\frac{(\mu_{{\rm pred},i} - \mu_i)^2}{2\sigma_{\mu,i}^2}\Bigr]}_{\rm distance} \;\times\; \underbrace{\bigl[1 + \delta_g(\mathbf{r})\bigr]}_{\substack{\rm density\\\rm weighting}},
\label{eq:prior}
\end{equation}
where $\mu_{{\rm pred},i} = 5\log_{10}(d_{L,i}/10\;{\rm pc})$ with $d_{L,i} = (1+z_{{\rm cos},i})\,r$, and $\sigma_{\mu,i}$ is the distance uncertainty for each SN. We note that systematic uncertainties are incorporated separately (see below). The $r^2$ term accounts for the increasing volume at larger distances, and the second term encodes the distance likelihood given the observed $\mu_i$. The final density term $[1+\delta_g(\mathbf{r})]$ evaluates the 3D overdensity field along the radial line of sight, upweighting regions where SNe are more likely to reside.
Figure~\ref{fig:los_posterior} illustrates how the density field reshapes $P(r)$ across different environments, from sharp unimodal peaks in overdense regions to broad multimodal features in voids. 

With flat priors on $\mathbf{V}_{\rm ext}$, $\beta$, and $\sigma_{v}$, the joint posterior over all SNe is then,
\begin{equation}
P(\mathbf{V}_{\rm ext}, \beta, \sigma_v \mid \{z_i\}) \propto \prod_{i}^{N} P(z_i \mid \beta, \mathbf{V}_{\rm ext}, \sigma_v).
\label{eq:joint_posterior}
\end{equation}

\subsection{Incorporating Systematic Covariance}
\label{sec:likelihood_sys}

Equation~\ref{eq:joint_posterior} treats each SN as statistically independent. However, systematic uncertainties such as photometric calibration, intrinsic scatter modeling, and selection effects introduce correlated shifts to SN distances that a product likelihood cannot capture. Typically, these correlations can be incorporated through a covariance matrix within a multivariate Gaussian likelihood \citep{2022Brout, Carreres_2025}, but applying this directly would replace the individual SN line of sight integrals with Gaussian approximations, losing the ability to account for inhomogeneous Malmquist bias.

Conceptually, our approach adds correlated Gaussian perturbations (from systematic uncertainties) to the non-Gaussian distance priors. Each systematic shifts the distance modulus linearly and has a Gaussian prior, so the systematic parameters can be marginalized analytically. This is equivalent to adding their covariance $\mathbf{C}_{\rm sys}$ to the likelihood \citep{Taylor_2010}. The covariance is constructed in $\mu$-space and is independent of the density field. The density field instead shapes the non-Gaussian priors, which we keep through numerical line-of-sight integration. The resulting hybrid log-likelihood is:
\begin{equation}
\ln\mathcal{L} = \sum_i \ln\mathcal{L}_{\mathrm{fwd},i} 
+ \Delta\ln\mathcal{L}_{\mathrm{sys}},
\label{eq:hybrid_lkl}
\end{equation}
where $\mathcal{L}_{\rm fwd,i}$ is the statistical-only forward likelihood (Equation~\ref{eq:los_integral}). Marginalizing over the line of sight contributes additional variance beyond $\sigma_v^2$ (Equation~\ref{eq:likelihood}), so we characterize each SN's velocity residual under $P(r)$ by its mean,
\begin{equation}
\bar{v}_i = \int P(r_i)\,\bigl[cz_i - cz_{{\rm pred},i}(r)\bigr]\,dr,
\label{eq:vbar}
\end{equation}
and line of sight variance,
\begin{equation}
\sigma^2_{{\rm LOS},i} = 
\int P(r_i)\,\bigl[cz_i - cz_{{\rm pred},i}(r)\bigr]^2\,dr - \bar{v}_i^2.
\label{eq:var_los}
\end{equation}
These moments are combined into an effective covariance matrix,
\begin{equation}
\mathbf{C}_{\rm eff} = \mathbf{C}_{{\rm sys},v} 
+ {\rm diag}(\sigma_v^2 + \sigma^2_{{\rm LOS},i}),
\label{eq:ceff}
\end{equation}
\noindent where $\sigma_v^{2}$ is from Equation~\ref{eq:likelihood}, $\sigma^2_{{\rm LOS},i}$ is the additional variance from marginalizing over distance, and $\mathbf{C}_{{\rm sys},v}$ is the systematic covariance transformed to velocity space. The systematic covariance term $\Delta\ln\mathcal{L}_{\mathrm{sys}}$ is the difference between a multivariate Gaussian over the $\bar{v}_i$ and its diagonal approximation, derived in Appendix~\ref{app:likelihood}:
\begin{equation}
\begin{split}
\Delta\ln\mathcal{L}_{\mathrm{sys}} = 
&\left[-\frac{1}{2}\left(\bar{\mathbf{v}}^{T} \mathbf{C}_{\rm eff}^{-1} 
\bar{\mathbf{v}} + \ln|\mathbf{C}_{\rm eff}|\right)\right] \\
- &\left[-\frac{1}{2}\sum_i\left(\frac{\bar{v}_i^2}{C_{{\rm eff},ii}} 
+ \ln C_{{\rm eff},ii}\right)\right].
\end{split}
\label{eq:sys_correction}
\end{equation}
This difference vanishes when the SNe are uncorrelated, recovering the \citet{Stahl_2021} likelihood. The diagonal of the systematic covariance is degenerate with $\sigma_v$, so only the correlations between SNe contribute.

Since $\sigma_{\mu,i}$ is already incorporated into $P(r_i)$ (Equation~\ref{eq:prior}), the systematic covariance is constructed in $\mu$-space to contain only the additional correlated uncertainties: $\mathbf{C}_{\rm sys} = \mathbf{C}_{\rm
tot} - \mathbf{C}_{\rm stat}$, where $\mathbf{C}_{\rm stat} = {\rm diag}(\sigma_{\mu,i}^2)$. To transform to velocity space we apply $\mathbf{C}_{{\rm sys},v} = \mathbf{J}\,\mathbf{C}_{\rm sys}\,\mathbf{J}^T$, where $\mathbf{J} = {\rm
diag}(J_i)$ and
\begin{equation}
J_i = \frac{d(cz)}{d\mu}\bigg|_{z_i} = \frac{\ln 10}{5} \cdot \frac{c\,d_L(z_i)}{d_C(z_i) + (1+z_i)\,c/H(z_i)},
\label{eq:jacobian}
\end{equation}
with $d_L$ and $d_C$ the luminosity and comoving distances respectively, and $H(z_i)$ the Hubble parameter at $z_i$.

We note that this conversion results in Gaussian errors in distance becoming log-normal in velocity space. However, the systematic shifts are small relative to each SN's statistical distance uncertainty, so the resulting covariance matrix provides an adequate description. 


\subsection{Systematic Uncertainty Analysis}\label{sec:systematics}

We examine the sensitivity of our $f\sigma_8$ measurement to common sources of uncertainty, following the framework established by previous SN\,Ia cosmological analyses \citep{Conley_2011,Betoule2014,Scolnic_2018,2022Brout}. We consider the following categories: observational biases from photometric calibration, measurement errors and foreground effects, SN modeling uncertainties arising from the standardization process, and reconstruction variants arising from selection effects in the redshift survey.

Each component shifts the measured SN distances, and these shifts build a covariance matrix,
\begin{equation}
C_{\rm sys}^{jk} = \sum_{n=1}^{N}\frac{\partial \mu_j}{\partial S_n}\frac{\partial \mu_k}{\partial S_n}\sigma^{2}_{\psi},
\label{eq:csys_construction}
\end{equation}
\noindent summed over all $N$ systematics, where $\partial\mu_j/\partial S_n$ is the shift in the $j$th distance modulus from the $n$th systematic $S_n$, while $\sigma_\psi$ is a weighting factor for each component. We transform $\mathbf{C}_{\rm sys}$ to velocity space following Equation~\ref{eq:jacobian}; this is the systematic covariance incorporated into the hybrid likelihood of Section~\ref{sec:likelihood_sys}.

We follow the DES 5YR analysis \citep{vincenzi2024dark} in our treatment of the SN observational and modeling systematics, with two exceptions: the survey-specific ATLAS calibration systematics, which are taken from \citet{Marlin_2025}, and the intrinsic scatter modeling, which follows the more conservative Pantheon$+$ approach \citep{2022Brout}. This also informs systematic weight choices for each component, as detailed below.

\begin{itemize}
\item \textbf{ATLAS photometric calibration}: The photometric calibration of the ATLAS survey was recently determined by \citet{Marlin_2025}, who measured average zeropoint (ZP) uncertainties of 5~mmag in the ``orange" band ($o$; approximately equal to a combined $r+i$ filter) and 10~mmag in the ``cyan" band ($c$; approximately $g+r$). They found central wavelength uncertainties that vary with each chip, but average 5\AA\ for $o$ and 57\AA\ for $c$.  We use the average uncertainties in this analysis, but when using real data we will use chip-by-chip systematic uncertainties for each of these terms.

\item \textbf{Calspec calibration}: We use zeropoint uncertainties for the Calspec system, from \citet{Bohlin20}, of 5~mmag per 7000\AA.

\item \textbf{Calibration of the SALT3 training data}: The DES 5YR analysis \citep{vincenzi2024dark} re-trained a sample of nine SALT models using random realizations of calibration uncertainties across each filter/telescope in the training data. These nine realizations are weighted in the systematic error covariance matrix to yield a $1\sigma$ shift in model training calibration.

\item \textbf{Redshift uncertainties}: Even small correlated uncertainties in measured redshifts can have a significant effect on cosmological parameter measurements.  We apply a global shift of $4 \times 10^{-5}$ in redshift to model this uncertainty following \citet[but see also \citealp{Calcino17}]{vincenzi2024dark}.  We note that this may be a conservative choice, as host-galaxy redshifts are substantially more precise than SN redshifts, and host redshifts are available for all SNe\,Ia in the TITAN sample on which we base our simulations.

\item \textbf{Milky Way extinction}: We incorporate systematic uncertainty arising from Milky Way (MW) dust attenuation by applying a 5$\%$ shift to the $E(B-V)$ map of \citet{Schlegel_1998}, following the recalibration by \citet{Schlafly_2011}.

\item \textbf{Milky Way reddening law}: We vary the MW reddening law by using a mean selective-to-total dust extinction parameter $R_V = 3.0$, which is a shift of 0.1 from the nominal Milky Way $R_V$, and we switch from the baseline reddening law of \citet{Fitzpatrick99} to the alternative reddening law of \citet{1989Cardelli}.

\end{itemize}

\subsubsection{SN Modeling Systematics}
We next account for systematic uncertainties in SN Ia modeling, and the resulting detection efficiency, including:

\begin{itemize}

\item \textbf{Intrinsic scatter models}:
The wavelength-dependence of SN\,Ia scatter about the Hubble diagram is a source of significant uncertainty.  
We use the BS21 model as our baseline, as it results in the best match to real data \citep{brout2021dust,Popovic_2023}, and adopt G10 and C11 as variants.  In the case of real data, these systematics primarily affect the determination of the Malmquist bias, and our systematic uncertainty budget therefore includes these scatter models by incorporating them as variants of the {\tt BiasCor} simulations (following previous cosmological analyses, e.g., \citealp{Jones19,2022Brout,vincenzi2024dark}).  However, we also produce cosmological parameter estimates using the G10 and C11 models as the basis of our ``data" simulations in Section \ref{sec:scatter}.

\item \textbf{Mass step correction}: We again follow Pantheon$+$ in allowing the host-galaxy mass step location (the divide between ``high" and ``low" host-galaxy masses) to shift by 0.2~dex to a value of $\log_{10}(M_{\ast}/M_{\odot}) = 10.2$, compared to our baseline mass step at $\log_{10}(M_{\ast}/M_{\odot}) = 10$.  We note that our baseline scatter model, BS21, explains the apparent mass step  by differences in the average reddening law between low-mass and high-mass hosts.  For the BS21 model, the mass step can be corrected for by either fitting for the mass step simultaneously with other nuisance parameters (as in the G10 and C11 scenarios), or by high-dimensional bias corrections.  Here, we use the former approach for computational efficiency.

\item \textbf{Selection effects}: To account for uncertainty in the ATLAS detection efficiency and the resulting Malmquist bias, we include a variant in which we linearly shift the measured efficiency curves by 0.2~mag in our {\tt BiasCor} simulation (i.e., the brightness at which 50\% of SNe are classified shifts by 0.2~mag).  This shift is primarily due to uncertainty in the ATLAS SN spectroscopic selection efficiency, and is the same size as the spectroscopic selection uncertainty in the DES five-year analysis \citep{Vincenzi_2021}.  Tweddle et al.\ (in prep.) will contain a more in depth discussion of the ATLAS selection efficiencies.

\end{itemize}

\begin{figure}
    \centering
    \includegraphics[width=0.9\linewidth]{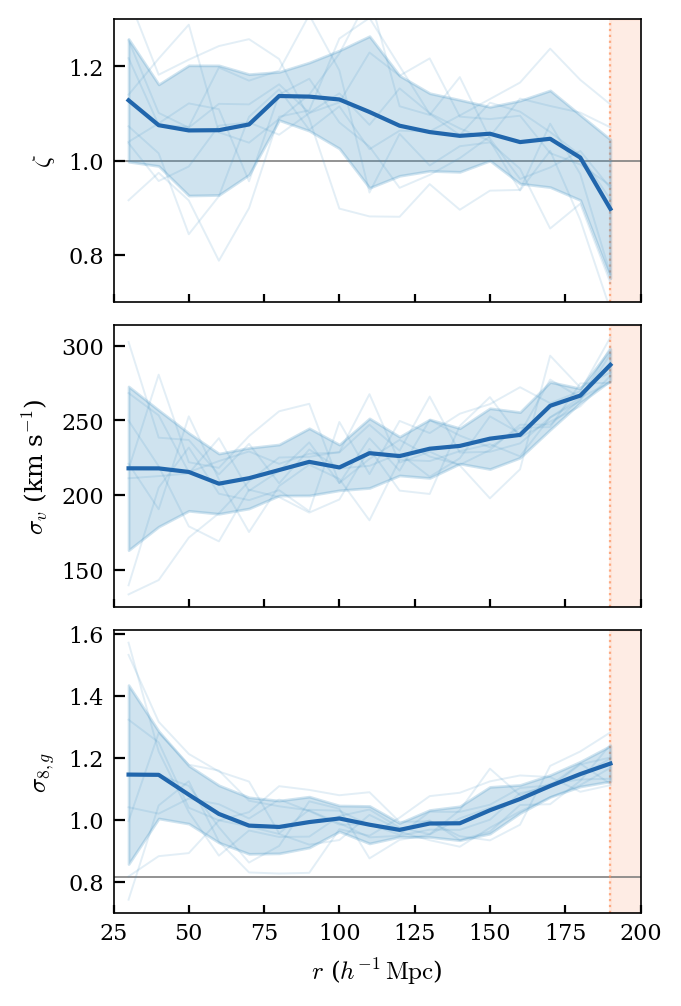}
    \caption{Radial diagnostics of the reconstruction quality across the 8 Uchuu mocks, showing variation in the regression slope $\zeta$, residual velocity dispersion $\sigma_v$, and $\sigma_{8,g}$, with increasing distance. Individual mocks are shown in gray with the ensemble mean in blue. The red shaded region marks $r > 190\,\Mpch$, beyond which all three quantities degrade due to boundary effects.} \label{fig:radial_diagnostics}
\end{figure}

\subsubsection{Density Reconstruction Systematics}\label{sec:recon_sys}

The density reconstruction systematics are treated separately from the SN covariance, since these variants alter the reconstructed field itself rather than introducing correlated shifts to SN distances. 

For each variant we repeat the analysis with the modified configuration and record the shift from baseline in each of the eight mocks, taking the mean shift as $\Delta f\sigma_8$ and the RMS of the per-mock shifts as the associated systematic uncertainty $\sigma_{\rm sys}$. Because the baseline and variant fits use identical SN samples, their shared statistical noise cancels in the difference, isolating the systematic effect. The total reconstruction systematic $\sigma_{\rm sys}^{\rm recon}$ is the quadrature sum of the three variant contributions.

\begin{itemize}
\item \textbf{Smoothing scale}: We adopt a Gaussian smoothing kernel of $R_s = 4\,\Mpch$ for our baseline reconstruction (Section~\ref{subsec:recon_results}), and examine variants at $R_s = 3$ and $R_s = 5\,\Mpch$. We take the RMS of the shifts from both variants, to characterize the impact as a single smoothing systematic.
\item \textbf{Galaxy survey depth}: The 2M$++$ catalog is flux-limited, introducing distance-dependent incompleteness and Malmquist bias in the reconstructed density field. To probe the effect of variations in the survey magnitude limit, we repeat our reconstruction and $f\sigma_8$ measurement with a relaxed cut of $K < 12.7$ (baseline $K < 12.5$).
\item \textbf{Boundary effects}: The reconstruction quality degrades near the edge of the survey volume (Figure~\ref{fig:radial_diagnostics}), and we evaluate $\sigma_{8,g}$ within $190\,\Mpch$ to avoid these effects in our baseline analysis. To quantify the impact of this choice, we re-evaluate $\sigma_{8,g}$ at $180\,\Mpch$ and propagate the change through to the $f\sigma_{8}$ measurement.
\end{itemize}

\section{Results}\label{sec:results}

\subsection{Baseline $f\sigma_8$ Recovery}

\begin{figure}
    \centering
    \includegraphics[width=1\linewidth]{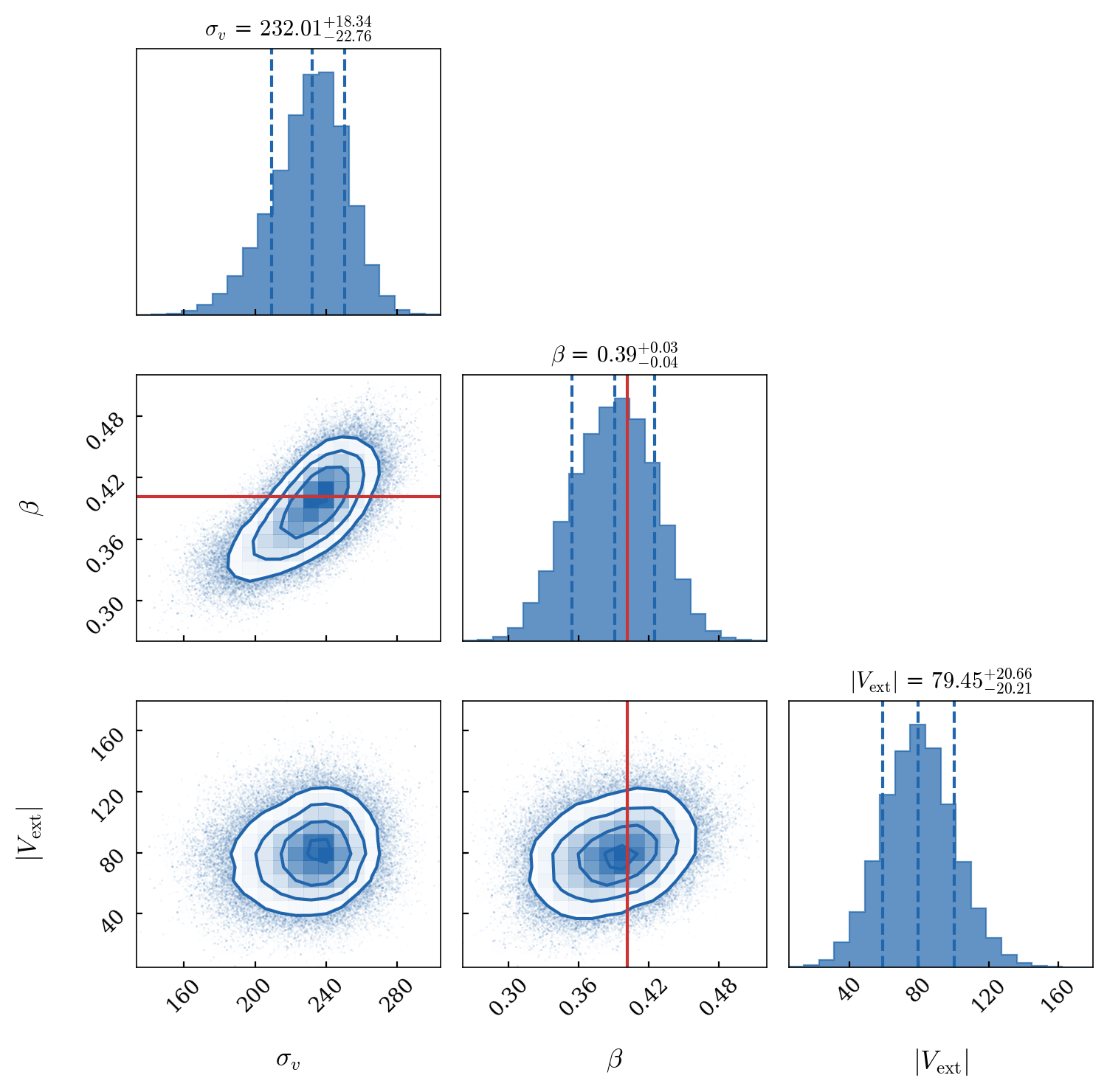}
    \caption{Posterior distributions from a single Uchuu mock for fitted parameters $\sigma_v$, $\beta$, and $|\mathbf{V}_{\rm ext}|$. The dashed line marks the median and the red line shows the simulated $\beta$ in this example.}
    \label{fig:corner}
\end{figure}

We measure $f\sigma_8$ from eight independent Uchuu mock realizations using the forward likelihood approach described in Section~\ref{sec:likelihood}. For each mock we sample the posterior using \texttt{emcee} \citep{MCMC_2013} with 5000 steps to determine $\beta$, $\mathbf{V}_{\rm ext}$, and $\sigma_v$. An example corner plot is shown in Figure~\ref{fig:corner}.

For the statistical-only baseline (Table~\ref{tab:fsigma8_validation}), we obtain $\langle f\sigma_8 \rangle = 0.429$ across the eight mocks (each consisting of between 2000 and 2077 SNe\,Ia). This result is a bias of just $0.1\%$ relative to the true simulated Uchuu value $f\sigma_8^{\rm Uchuu} = 0.428$. The mean per-mock statistical uncertainty is $\sigma_{\rm stat} = 0.030$, consistent with the overall mock-to-mock scatter of $0.031$, indicating that our error estimates are well calibrated. Our observed scatter is $7\%$ of $f\sigma_8$, slightly larger than the $\sim5\%$ cosmic variance estimated by \citet{Hollinger24} for the reconstruction alone, which we attribute to additional scatter from SN Ia distance uncertainties and the forward likelihood marginalization.

\begin{table}
    \centering
    \caption{Statistical-only baseline recovery of $f\sigma_8$ from the eight independent Uchuu mocks. The true value $f\sigma_8^{\rm Uchuu} = 0.428$ is fixed by the Uchuu cosmology \citep{Planck_2015} and is common to all mocks. The mock-to-mock scatter is $0.031$.}
    \label{tab:fsigma8_validation}
    \begin{tabular}{@{}lccc@{}}
        \toprule
        Mock & $N_{\rm SNe}$ & $f\sigma_8$ & $\sigma_{\rm stat}$ \\
        \midrule
        0 & 2019 & $0.450$ & $0.021$ \\
        1 & 2000 & $0.417$ & $0.038$ \\
        2 & 2077 & $0.478$ & $0.031$ \\
        3 & 2049 & $0.389$ & $0.027$ \\
        4 & 2007 & $0.428$ & $0.035$ \\
        5 & 2049 & $0.384$ & $0.034$ \\
        6 & 2035 & $0.441$ & $0.030$ \\
        7 & 2074 & $0.441$ & $0.027$ \\
        \midrule
        Mean &  & $0.429$ & $0.030$ \\
        \bottomrule
    \end{tabular}
\end{table}

\begin{figure}
    \centering
    \includegraphics[width=0.9\linewidth]{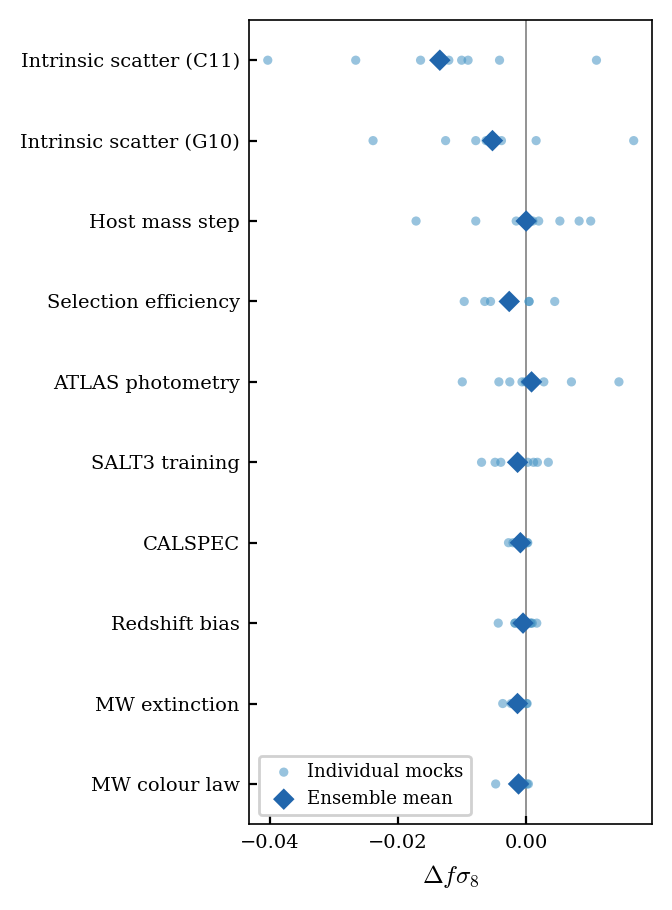}
    \caption{Shift in $f\sigma_8$ for each individual SN systematic relative to the statistical-only baseline. The blue circles represent the individual mock realizations, while the diamonds show the ensemble mean. The most significant shifts across the mocks are aligned with the intrinsic scatter model variations. Reconstruction systematics are reported separately in Table~\ref{tab:sys_budget}.}
    \label{fig:systematic_shifts}
\end{figure}

\subsection{Systematic Error Budget}
\label{sec:sys_budget}

When a systematic covariance term is added to $\mathbf{C}_{\rm eff}$ (Equation~\ref{eq:ceff}), the resulting fit differs from the statistical-only baseline in both the best-fit value and posterior width. For each SN systematic described in Section~\ref{sec:systematics}, we obtain two quantities: (i) the mean shift in best-fit value, $\Delta f\sigma_8 = \langle f\sigma_{8,i} - f\sigma_{8,\rm stat} \rangle$, and (ii) the additional systematic uncertainty,
\begin{equation}
\sigma_{{\rm sys},i} = \sqrt{\sigma_{{\rm tot},i}^2 - \sigma_{\rm stat}^2},
\label{eq:sigma_sys}
\end{equation}
where $\sigma_{{\rm tot},i}$ is the posterior width when systematic $i$ is included in $\mathbf{C}_{\rm eff}$, and $\sigma_{\rm stat}$ is the corresponding statistical-only error; both are averaged over the eight mock realizations. The reconstruction systematics (Section~\ref{sec:recon_sys}) do not enter $\mathbf{C}_{\rm eff}$. Rather than broadening the posterior, each variant shifts the recovered $f\sigma_8$, and we take $\sigma_{{\rm sys},i}$ as the RMS of these per-mock shifts.

Figure~\ref{fig:systematic_shifts} shows $\Delta f\sigma_{8}$ for each systematic, for both individual mocks and the ensemble mean, and Table~\ref{tab:sys_budget} summarizes the mean shift and additional uncertainty for each component. Among the SN systematics, the intrinsic scatter models produce the largest shifts. For the C11 model we determine $\Delta f\sigma_8 = -0.014 \pm 0.007$, for G10 $\Delta f\sigma_8 = -0.005 \pm 0.005$, where both shift the best-fit value toward lower $f\sigma_8$ values. The remaining calibration, Milky Way, and survey selection terms each shift by $|\Delta f\sigma_8| \leq 0.003$. The posterior-broadening contributions ($\sigma_{\rm sys}$) are largest for the intrinsic scatter models (0.007--0.009), host mass step (0.006), and selection efficiency (0.006), with all calibration and Milky Way terms $\leq 0.005$. 

For the reconstruction systematics (Section~\ref{sec:recon_sys}), the smoothing scale shifts $f\sigma_8$ by $0.002 \pm 0.012$, the galaxy survey depth by $0.002 \pm 0.007$, and the boundary choice by $-0.008 \pm 0.008$, effectively a decrease when examined within a smaller volume. We note that none of these shifts are statistically significant, but do add scatter to the measurements.

Combining all the SN systematics in quadrature yields $\sigma_{\rm sys}^{\rm SN} = 0.017$, while the reconstruction variants contribute $\sigma_{\rm sys}^{\rm recon} = 0.016$, for a total systematic uncertainty of $\sigma_{\rm sys} = 0.023$. Adding this in quadrature to the statistical uncertainty ($\sigma_{\rm stat} = 0.030$) gives a total uncertainty of $\sigma_{\rm tot} = 0.038$. This quadrature sum assumes the components are independent and is therefore likely conservative in our error budget.

\begin{deluxetable*}{llllcc}
\tablecaption{Systematic uncertainty breakdown for $\fsigma$, averaged over eight mock realizations (Section~\ref{sec:sys_budget}). Each SN systematic is applied as the variation $S_\psi$ weighted by $\sigma_\psi$; $\sigma_{\rm sys}$ is the resulting posterior broadening and $\Delta\fsigma$ is the mean shift in the best-fit value, both relative to the baseline. For the reconstruction systematics, $\sigma_{\rm sys}$ is instead the RMS of the per-mock shifts. The SN and reconstruction contributions are combined in quadrature as $\sigma_{\rm sys}^{\rm SN}$ and $\sigma_{\rm sys}^{\rm recon}$, and the total uncertainty is $\sigma_{\rm tot} = \sqrt{\sigma_{\rm stat}^2 + (\sigma_{\rm sys}^{\rm SN})^2 + (\sigma_{\rm sys}^{\rm recon})^2}$.}
\label{tab:sys_budget}
\tablehead{
\colhead{\textbf{Component}} & \colhead{\textbf{Baseline}} & \colhead{\textbf{Systematic ($S_\psi$)}} & \colhead{\textbf{$\sigma_\psi$}} & \colhead{\textbf{$\sigma_{\rm sys}$}} & \colhead{\textbf{$\Delta\fsigma$}}
}
\startdata
\underline{Calibration} & & & & & \\
Photometry & Nominal ZP \& $\lambda_{\rm eff}$ & ZP: $c\pm10$, $o\pm5$\,mmag; $\lambda_{\rm eff}$: $c\pm57$, $o\pm5$\,\AA & 1 & $0.003$ & $+0.001$ \\
CALSPEC & Nominal spectrophotometry & $\pm7.1$\,mmag/$z$ slope & 1 & $0.004$ & $-0.001$ \\
SALT3 training & SALT3.K21 & 9 surface variants & $1/3$ & $0.002$ & $-0.001$ \\
\\
\underline{Survey} & & & & & \\
Redshift bias & No $z$-shift & $10^{-4}$ $z$-shift & 0.4 & $0.005$ & $-0.001$ \\
Selection efficiency & Nominal & $+0.2$\,mag shift & 1 & $0.006$ & $-0.003$ \\
\\
\underline{SN Modeling} & & & & & \\
Host mass step & Split at 10.0 & Split at 10.2 & 1 & $0.006$ & $+0.000$ \\
Intrinsic scatter (G10) & BS21 bias correction & G10 bias correction & 0.7 & $0.009$ & $-0.005$ \\
Intrinsic scatter (C11) & `` '' & C11 bias correction & 0.7 & $0.007$ & $-0.014$ \\
\\
\underline{Milky Way} & & & & & \\
$E(B{-}V)$ scale & \citet{Schlafly_2011} & $\Delta E(B{-}V) = 5\%$ & 1.0 & $0.005$ & $-0.001$ \\
Reddening law & \citet{Fitzpatrick99} & $\Delta R_{V} = 0.1$ & 0.3 & $0.004$ & $-0.001$ \\
\\
\underline{Reconstruction} & & & & & \\
Smoothing & $R_s=4$ & $R_s=3,\,5$ & --- & $0.012$ & $+0.002$ \\
Galaxy survey depth & $K<12.5$ & $K<12.7$ & --- & $0.007$ & $+0.002$ \\
Boundary & $R_{\rm max}=190$\,\Mpch & $R_{\rm max}=180$\,\Mpch & --- & $0.008$ & $-0.008$ \\
\hline
& & & $\sigma_{\rm stat}$ & $0.030$ & \\
& & & $\sigma_{\rm sys}^{\rm SN}$ & $0.017$ & \\
& & & $\sigma_{\rm sys}^{\rm recon}$ & $0.016$ & \\
& & & $\bm{\sigma_{\rm tot}}$ & $\bm{0.038}$ & \\
\hline
\enddata
\end{deluxetable*}

\subsection{Sensitivity to the Intrinsic Scatter Model}
\label{sec:scatter}

\begin{figure}
    \centering
    \includegraphics[width=1\linewidth]{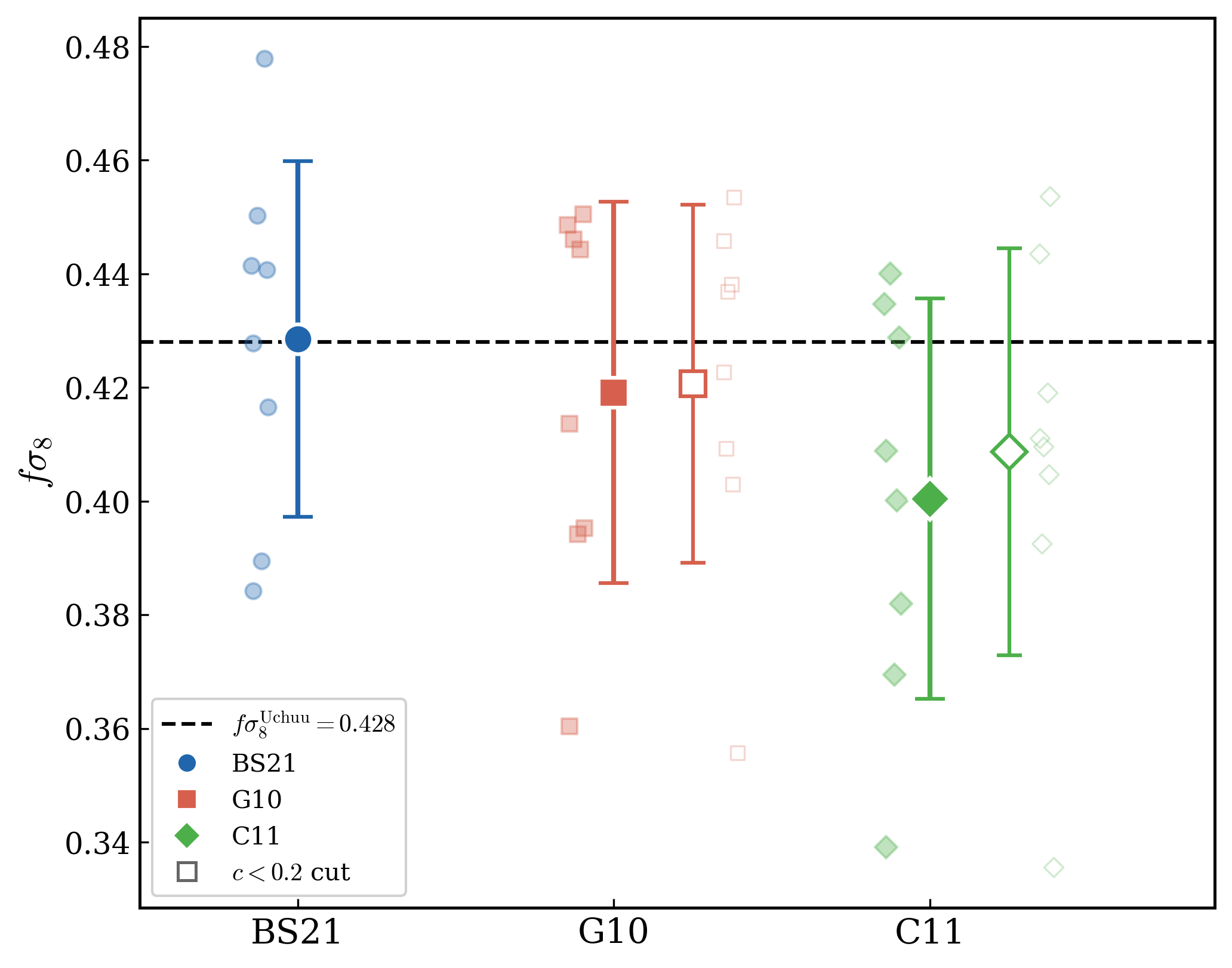}
    \caption{Recovery of $f\sigma_8$ across the three intrinsic scatter models (BS21, G10, C11) from 8 Uchuu mocks. Small symbols show individual mocks, while the larger symbols show the ensemble mean with mock-to-mock scatter as the error bars. Open symbols show the impact of a $c < 0.2$ cut applied at the {\tt BiasCor} stage, while the dashed line highlights $f\sigma_8^{\rm Uchuu} = 0.428$.}
    \label{fig:scatter_models}
\end{figure}

While we use BS21 for our baseline analysis, we additionally simulate our ``data'' sample for each mock with both the G10 and C11 intrinsic scatter models, allowing us to examine the sensitivity of our recovered $f\sigma_8$ values to this choice. In Figure~\ref{fig:scatter_models}, we compare $f\sigma_8$ measurements for each scatter model, finding that with our baseline model we recover the true value to within $0.1\%$. For the G10 and C11 cases, we find ensemble means of $0.419$ ($-2.1\%$, $0.7\sigma$) and $0.400$ ($-6.4\%$, $2.2\sigma$) respectively, where the significance of each shift is taken relative to the standard error of the mean. These offsets differ from the intrinsic scatter covariances in Table~\ref{tab:sys_budget}, which hold the ``data'' fixed at BS21 and instead vary the scatter model used in the bias correction. Here each scatter model has its own data simulation, so the resulting offsets are larger and we find mock-to-mock scatter is comparable across all three ($\sigma = 0.031$, $0.034$, $0.035$).

While the discrepancy in the C11 model is concerning, we note that in the recent cosmology analysis from the Dark Energy Survey (DES), \citet{vincenzi2024dark} find that the G10 and C11 scatter models provide a notably worse fit to the SN\,Ia data and these models are not included in their systematic error budget (but were included in Pantheon$+$).  It is therefore likely that both the G10/C11 bias correction systematics, and the simulated C11 results presented here, are overly conservative assessments of the likely bias in the real data.  

In examining the C11 simulations, we find that the likely cause of the C11 bias is that the simulated populations for both C11 and G10 favor an extended red tail in color compared to the BS21 population.  While G10 results are largely insensitive to redder colors, C11 predicts a color distribution that becomes intrinsically more skewed at redder colors, mimicking the effect of gravitationally induced velocities (see also \citealp{Carreres_2025}). Applying a $c < 0.2$ cut before the {\tt BiasCor} stage reduces the C11 offset from $-6.4\%$ to $-4.6\%$ (1.6$\sigma$). When these simulations are refined for the real data, the comparisons uncovered here will inform color cuts on the data that limit intrinsic skewness and ultimately improve $f\sigma_{8}$ estimation.


Lastly, while we rely on a one-dimensional, redshift dependent bias correction in this work, bias corrections that correct for selection effects in the $x_1$ and $c$ parameters have been shown to increase precision in Hubble residuals and cosmological parameter inference.  It is likely that these higher-dimensional corrections will reduce the apparent bias from the C11 model by correcting Hubble residual trends as a function of $x_1$ and $c$ \citep[e.g.,][]{2016Scolnic}.


\section{Discussion}\label{sec:discussion}

This work builds upon the reconstruction-and-scaling methodology of \citet{Stahl_2021} by incorporating SN systematics into SN\,Ia $f\sigma_8$ measurements for the first time. Using eight Uchuu mock realizations, we demonstrate that this methodology is robust across independent density fields. Our analysis validates the method and provides the first systematic uncertainty budget for this type of measurement, in preparation for the upcoming TITAN SN\,Ia analysis (Dixon et al., in prep.). We next discuss the main sources of systematic uncertainty in our method, and then discuss the implications for current and future measurements.

\subsection{Systematic uncertainties in the reconstruction-and-scaling method}

Recent studies have highlighted the importance of a comprehensive covariance treatment for robust $f\sigma_8$ measurements. \citet{Blake24} found that the reconstruction-and-scaling method underestimates uncertainties in $\beta$ when the full covariance of the reconstructed velocity field is not accounted for. Additionally, \citet{Turner_2023} used idealized simulations to find that the reconstruction-and-scaling method can produce biased $f\sigma_8$ parameter estimation.  Our work, however, does not find evidence for either effect, with statistically insignificant $f\sigma_8$ bias and mock-to-mock scatter that is consistent with our derived uncertainties.  We suspect that the most important factors that may differ in our analysis are 1) the tuning of the smoothing scale to reduce the effects of nonlinearities while not over-smoothing the density fluctuations, and 2) a sufficiently large galaxy catalog that samples the density field to scales well below 4 \Mpch.

The choice of SN Ia intrinsic scatter model remains one of the main sources of systematic error in our analysis. In Section~\ref{sec:scatter}, we found that using the C11 model in place of BS21 results in a bias of $-0.028$ ($-6.4\%$). Examining this further, we found this is partly driven by redder SNe with a skewed distribution, and after implementing a color cut $c<0.2$, reduces this bias to $-0.019$ ($-4.6\%$). Using instead a maximum likelihood PV method, \citet{Carreres_2025} uncovered larger biases (up to $\sim$$-20\%$) for the BS21 model, driven by non-Gaussian distribution of Hubble diagram residual, highlighting that intrinsic scatter model choice affects measurements of $f\sigma_8$ across different approaches, and that ultimately it is important to understand these effects. 

Two additional improvements in future work may help our analysis reduce statistical and systematic errors further. The first is the use of higher dimensional bias corrections \citep{Popovic_2021}, which increase precision by correcting Hubble residual trends introduced by selection effects as a function of $x_1$, $c$, $\alpha_{\rm SN}$, $\beta_{\rm SN}$ and redshift.  The second is the ability to simultaneously marginalize over SN\,Ia nuisance parameters alongside $f\sigma_8$; previous work from \citet{Stahl_2021} included this option in their forward likelihood model, but in this analysis we instead use the BBC approach to estimate these parameters from large Monte Carlo SN\,Ia simulations over an $\alpha$ and $\beta$ grid, which corrects for observational biases.  Although the BBC method simultaneously estimates distances binned in redshift space, which is appropriate for, e.g., $H_0$ measurements, for $f\sigma_8$ it may be more accurate to simultaneously measure binned distances as a function of e.g., the galaxy overdensity, which we will explore in future work.

Regarding additional reconstruction systematics, our mocks exhibit similar radial degradations (Figure~\ref{fig:radial_diagnostics}) in $\zeta$, $\sigma_v$, and $\sigma_{8,g}$ to those found by \citet{Hollinger24} in their 2M$++$ mock density fields. This motivates us to evaluate $\sigma_{8,g}$ within $190\,\Mpch$, as the effects become more significant at larger distances, especially near the reconstruction boundary. This conservative cut avoids the edge effects (within $20\,h^{-1}\mathrm{Mpc}$ of the survey boundary) reported by \citet{Hollinger24}, where the velocity scatter increases by $\sim75\,\kms$ and the fitted $\beta$ can be biased by 20--30\%. While this systematic has less impact in our analysis, we include it as a variant by instead evaluating $\sigma_{8,g}$ within $180\,\Mpch$. Other 2M$++$ reconstruction systematics, such as the Zone of Avoidance mask (where the Milky Way's dust and stars obscure distant galaxies) and incomplete sky coverage, are not explored here, as our mocks are built from full Uchuu subvolumes without a Galactic sky mask. They therefore mimic rather than replicate the 2M$++$ angular selection, since our goal is to test how systematic covariance affects $f\sigma_8$, not to reproduce the full survey geometry. However, \citet{Hollinger24} have examined these effects, finding that survey selection collectively biases $f\sigma_8$ high by a factor of ${\sim}1.04$.

\subsection{Implications for current and future $f\sigma_8$ measurements}\label{sec:comparison}

Our baseline analysis gives $\langle f\sigma_8 \rangle = 0.429 \pm 0.038$ ($\sigma_{\rm stat}=0.030$, $\sigma_{\rm sys}=0.023$), consistent with the Planck 2015 cosmology used in the Uchuu simulation ($f\sigma_8^{\rm Uchuu}=0.428$) to within $0.1\%$. This shows our reconstruction-and-scaling method recovers the true growth rate while incorporating a full systematic covariance budget.

Previous reconstruction-and-scaling analyses using 2M$++$ considered only statistical errors with no systematic budget, reporting $f\sigma_8 = 0.40 \pm 0.07$ \citep{Turnbull_2012}, $0.400 \pm 0.017$ \citep{Boruah_2020}, and $0.390 \pm 0.022$ \citep{Stahl_2021}. Our statistical uncertainty exceeds the latter two despite our larger sample size. The likely factors driving this difference between the simulated TITAN sample and these compilations include the redshift distribution and light-curve standardization, rather than the method itself. We leave a more direct comparison of uncertainties to the upcoming real-data analysis (Dixon et al., in prep), where the same 2M$++$ reconstruction and the TITAN SNe\,Ia will allow a like-for-like comparison.

Our systematic error budget ($\sigma_{\rm sys}=0.023$) is the first to account for both SN and reconstruction systematics, with the reconstruction contribution alone ($\sigma_{\rm sys}^{\rm recon}=0.016$) comparable to the statistical errors of those analyses. Although the exact budget depends on each analysis, adding a systematic covariance of this size would increase their total uncertainties by $\sim45\%$ (e.g., $0.022$ to $\approx0.032$ for \citealt{Stahl_2021}), split roughly evenly between reconstruction and SN systematics. The fractional increase is largest for analyses with small statistical errors. For our own analysis the increase is $\sim25\%$ ($0.030$ to $0.038$). Statistical-only error bars therefore under-represent the true uncertainty, and the modest tensions with $\Lambda$CDM reported in these works (e.g., $\sim1.4\sigma$ in \citealt{Stahl_2021}) may become insignificant once systematics are included.

The sensitivity to intrinsic scatter also affects the interpretation of previous measurements. As shown in Section~\ref{sec:scatter}, using the C11 model instead of our baseline BS21 biases the recovered $f\sigma_8$ by $-6.4\%$, moving the result further from the Planck $\Lambda$CDM value ($f\sigma_8 \approx 0.43$). This bias is partly driven by a redder color distribution, and after applying a $c<0.2$ cut the bias reduces to $-4.6\%$, albeit still a $1.6\sigma$ effect. Thus, if the real SN\,Ia sample contains red SNe with a skewed color distribution, an incorrect scatter model can bias $f\sigma_8$ low and increase the apparent tension. However, our baseline BS21 model, which provides a better fit to the Hubble diagram residuals, recovers the simulated value with negligible bias. This effect could also have increased the apparent tension in previous measurements, particularly depending on the $c$ distribution of the SN samples, but we show it can be mitigated in this analysis.

In our analysis we are limited to $z<0.067$ by the 2M$++$ reconstruction volume, but the full TITAN sample contains $>3000$ SNe\,Ia at $z<0.1$. Extending the analysis to this larger volume would reduce cosmic variance and enable a more detailed study of the redshift evolution of the growth rate. Combining with other SN catalogs (e.g., Pantheon$+$) following the approach of \citet{Stahl_2021} would also extend sky coverage and probe across a wider range of environments. Alternative density reconstructions such as Manticore \citep{McAlpine_2025}, CosmicFlows4 (CF4; \citealt{Tully_2023}), and CF4$++$, which incorporates data from WALLABY, FAST, and DESI \citep{Courtois_2025, Hollinger_2026}, offer the potential to extend this methodology to $z \simeq 0.1$ and provide valuable systematic cross-checks. A larger volume also has the benefit of more accurately measuring $\sigma_{8,g}$ \citep{Hollinger24}.

We will detail a more direct comparison of SN-based $f\sigma_8$ measurements with Planck and other large-scale structure probes in the upcoming TITAN paper (Dixon et al., in prep), where the real SN\,Ia sample will be analyzed with the full systematic covariance framework validated here. This analysis aims to quantify the expected scale and impact of these contributions and provide a road map for future measurements.


\section{Conclusions}\label{sec:conclusions}

We present a simulated end-to-end measurement of $f\sigma_{8}$ using mock ATLAS SN\,Ia catalogs generated from the Uchuu $N$-body simulations, validating the reconstruction-and-scaling approach and characterizing its systematic uncertainty budget for the first time. To incorporate systematic uncertainties while preserving the non-Gaussian distance priors from the density field, we develop and apply a hybrid likelihood that combines the forward likelihood framework of \citet{Stahl_2021} with a full systematic covariance.

From eight independent mock realizations, each containing ${\sim}2000$ SNe\,Ia at $z < 0.067$, we recover $\langle f\sigma_8 \rangle = 0.429 \pm 0.038$ ($\sigma_{\rm stat} = 0.030$, $\sigma_{\rm sys} = 0.023$), consistent with the Uchuu input value ($f\sigma_8 = 0.428$) to within $0.1\%$. The systematic contributions are similar in scale, from SN\,Ia ($\sigma_{\rm sys}^{\rm SN}=0.017$) and the density reconstruction ($\sigma_{\rm sys}^{\rm recon}=0.016$). The mock-to-mock scatter of $0.031$ is consistent with the overall statistical uncertainty, confirming that our error estimates are well calibrated.

Comparing to earlier reconstruction-and-scaling analyses, we find that adding a comparable systematic budget would increase their total uncertainties by $\sim45\%$. Statistical-only error bars may therefore understate the true uncertainty, and the modest tensions with $\Lambda$CDM reported in some of those works may soften once systematics are included. We also find that the assumed intrinsic scatter model, and the resulting bias corrections, can shift $f\sigma_8$ to lower values; this appears largely driven by non-Gaussian, color-related effects, and can artificially exacerbate tension with $\Lambda$CDM. However, our baseline BS21 model recovers the true value with negligible bias.

The framework and systematic budget presented here provides a foundation for future SN\,Ia peculiar velocity measurements of $f\sigma_8$. As statistical uncertainties decrease with larger SN\,Ia samples, constraining systematics, especially in the low-$z$ regime, is vital to a complete picture of the different contributions. The upcoming TITAN analysis (Dixon et al., in prep.) will apply this methodology to real data and assess the result in the context of other low-$z$ probes and the wider $f\sigma_8$ tension.

\begin{acknowledgments}
We would like to thank Dan Scolnic for useful discussions.

M.D. and D.O.J. acknowledge support from NSF grants AST-2407632, AST-2429450, and AST-2510993, NASA grants 80NSSC24M0023 and 80NSSC24K0353, and HST/JWST grants HST-GO-17128.028 and JWST-GO-05324.031, awarded by the Space Telescope Science Institute (STScI), which is operated by the Association of Universities for Research in Astronomy, Inc., for NASA, under contract NAS5-26555.  This work is also funded by the Gordon and Betty Moore Foundation through Grant GBMF13900 to D.O.J.

SJS  and KWS acknowledge funding from STFC Grants ST/Y001605/1, ST/X001253/1, a Royal Society Research Professorship and the Hintze Family Charitable Foundation. 

AD is supported by the European Union’s Horizon 2020 research and innovation programme under ERC Grant Agreement No. 101002652 (BayeSN; PI K. Mandel) and Marie Skłodowska-Curie Grant Agreement No. 873089 (ASTROSTAT-II).

This work used resources of the National Energy Research Scientific Computing Center (NERSC).
N.E.D acknowledges support from NSF grants LEAPS-2532703 and AST-2510993.

This work has made use of data from the Asteroid Terrestrial-impact Last Alert System (ATLAS) project. ATLAS is primarily funded to search  for near earth asteroids through NASA grants NN12AR55G, 80NSSC18K0284,  and 80NSSC18K1575; byproducts of the NEO search include images and  catalogs from the survey area.  The ATLAS science products have been  made possible through the contributions of the University of Hawaii  Institute for Astronomy, the Queen's University Belfast, the Space  Telescope Science Institute, the South African Astronomical Observatory (SAAO),  and the Millennium Institute of Astrophysics (Chile) the Instituto de Astrofisica de Canarias (Spain) and the University of Oxford. 

\end{acknowledgments}

\bibliographystyle{aasjournal}
\bibliography{ref}

\appendix
 
\section{Derivation of the Hybrid Log-Likelihood}
\label{app:likelihood}

In Section~\ref{sec:likelihood_sys}, we introduced a hybrid log-likelihood (Equation~\ref{eq:hybrid_lkl}) that adds a systematic covariance term $\Delta\ln\mathcal{L}_{\mathrm{sys}}$ to the SN forward likelihood framework. Here we show the derivation leading to Equation~\eqref{eq:sys_correction}. The forward likelihood (Equations~\ref{eq:likelihood}--\ref{eq:prior}) keeps the full non-Gaussian distance prior $P(r)$ for each SN, but treats the SNe as statistically independent. To describe correlations between SNe, we summarize each SN's velocity residual under $P(r)$ by its mean $\bar{v}_i$ and variance $\sigma^2_{{\rm LOS},i}$ (Equations~\ref{eq:vbar}--\ref{eq:var_los}). The multivariate Gaussian log-likelihood for the residual vector $\bar{\mathbf{v}}$ with covariance $\mathbf{C}_{\rm eff}$ (Equation~\ref{eq:ceff}) is then
\begin{equation}
\ln\mathcal{L}_{\rm cov} = -\frac{1}{2}\bigl[\bar{\mathbf{v}}^{T} \mathbf{C}_{\rm eff}^{-1}\bar{\mathbf{v}} + \ln|\mathbf{C}_{\rm eff}| + N\ln 2\pi\bigr].
\label{eq:app_Lcov}
\end{equation}
This captures the correlated uncertainties but replaces the non-Gaussian structure of $P(r)$. The forward likelihood already describes each individual SN, so we only want the part of $\ln\mathcal{L}_{\rm cov}$ that describes the correlations between SNe. We therefore subtract its diagonal approximation, which treats each SN independently with variance $C_{{\rm eff},ii} = \sigma_v^2 + \sigma^2_{{\rm LOS},i} + C_{{\rm sys},v,ii}$,
\begin{equation}
\ln\mathcal{L}_{\rm diag} = -\frac{1}{2}\sum_i \!\left[\frac{\bar{v}_i^2}{C_{\rm eff,ii}} + \ln(2\pi C_{\rm eff,ii})\right],
\label{eq:app_Ldiag}
\end{equation}
and avoid double-counting the variance captured by the forward likelihood. The hybrid log-likelihood is then
\begin{equation}
\ln\mathcal{L} = \underbrace{\sum_i \ln\mathcal{L}_{\mathrm{fwd},i}}_{\text{statistical-only}} 
\;+\; \underbrace{\bigl(\ln\mathcal{L}_{\mathrm{cov}} 
- \ln\mathcal{L}_{\mathrm{diag}}\bigr)}_{\Delta\ln\mathcal{L}_{\mathrm{sys}}}.
\end{equation}
The $N\ln 2\pi$ terms cancel in the difference, reducing to Equation~\ref{eq:sys_correction}. This difference depends only on the off-diagonal correlations of $\mathbf{C}_{\rm eff}$: when $\mathbf{C}_{{\rm sys},v}$ is purely diagonal, $\ln\mathcal{L}_{\rm cov} = \ln\mathcal{L}_{\rm diag}$, the correction vanishes, and the likelihood reduces to the independent statistical-only form (Equation~\ref{eq:joint_posterior}).

\end{document}